\documentclass[onecolumn,tightenlines,nofootinbib,preprintnumbers,article,
amsmath,amssymb]{revtex4}

\usepackage[usenames,dvipsnames]{color}
\usepackage{caption2}
\usepackage{dcolumn}
\usepackage{graphicx}
\usepackage{subfigure}
\usepackage{epstopdf}
\usepackage{bm}
\usepackage{lscape}
\usepackage{picture}
\usepackage{longtable}
\usepackage{supertabular}
\usepackage{array}
\usepackage{tabularx}
\usepackage{amsmath}
\usepackage{siunitx}
\usepackage{booktabs}
\renewcommand{\arraystretch}{0.9}
\usepackage{setspace}

\begin{document}
\begin{spacing}{1.5}

\title{The $CP$ violations  and branching ratios for $B_c^+\to D_{(s)}^+\pi^+\pi^-(K^{+}K^{-})$ from  interference of the vector mesons in Perturbative QCD}

\author{Kun Shuai Ye$^{1}$\footnote{Email: 1620446906@qq.com}, Gang L\"{u} $^{1}$\footnote{Email: ganglv66@sina.com},
Na-Wang$^{1}$\footnote{Email: wangna@haut.edu.can},  Jian Chai$^{1}$\footnote{Email:physchai@hnu.edu.cn},
Xin-Heng Guo $^{2}$\footnote{Email: xhguo@bnu.edu.cn}}

\affiliation{\small $^{1}$ School of Physics, Henan University of Technology, Zhengzhou 450001, China\\
	\small $^{2}$College of Nuclear Science and Technology, Beijing Normal University, Beijing 100875, China\\}

\begin{abstract}
Within the framework of the perturbative QCD approach utilizing $K_T$ factorization, we have investigated the CP violations and branching ratios in the decay processes of $B_{c}^{+}\to D_{(s)} ^{+}V(V\rightarrow\pi^{+}\pi^{-})$ and $B_{c}^{+}\to D_{(s)}^{+}V(V\rightarrow K^{+}K^{-})$, where V denotes three vector mesons $\rho^0$, $\omega$, and $\phi$. During the $V\to \pi^+\pi^-$ and $V\to K^+K^-$ decay processes, we incorporated the $\rho^{0}-\omega-\phi$ mixing mechanism to describe the amplitudes of these quasi-two-body decay processes. Within the interference regime of the three vector particles, we observed distinct changes in both CP violations and branching ratios. Furthermore, our study presents evidence for local CP violations and branching ratios that warrants further investigation through experiments.
\end{abstract}

\maketitle

\section{Introduction}
\label{sec:sample1}
The $B_{c}$ meson stands out due to its composition of two heavy quarks, which makes it the flavor-asymmetric heavy meson with such a configuration. It is initially observed by the CDF collaboration in 1998 \cite{CDF:1998axz}. In the context of the Standard Model (SM), this meson possesses definite b and c quantum numbers and lies below the threshold for BD meson formation \cite{Nayak:2022qaq}. Consequently, it remains stable against strong and electromagnetic interactions, undergoing only weak decay processes \cite{Gouz:2002kk}.
The weak decay of the $B_{c}$ meson is typically categorized into three types due to the comparable contributions from both heavy quarks that constitute it: (1) $b$ quark decay with the $c$ quark as a spectator; (2) $b$ quark as a spectator with the $c$ quark decay; and (3) annihilation of both the $b$ and $c$ quarks, which contributes less significantly. Estimates suggest that approximately $70\%$ of the total decay rate arises from c-quark decays, while b-quark decays and annihilation contribute $20\%$ and $10\%$, respectively \cite{Zou:2017yxc}. The diverse range of weak decay channels offered by the $B_{c}$ meson presents an opportunity to test the Standard Model and search for potential signals of new physics \cite{Brambilla:2010cs}.

Currently, various methodologies are employed to investigate the weak decay of the $B_{c}$ meson, including QCD factorization (QCDF) \cite{Beneke:2003zv,Beneke:1999br}, perturbative QCD (PQCD) \cite{Lu:2013jma,Lu:2016lgc}, and the soft-collinear effective theory (SCET) \cite{Bauer:2001yt}. 
 The $B_{c}$ meson is a non-relativistic heavy quark system, implying that both quarks in the $B_{c}$ meson are stationary and non-relativistic. Since the charm quark in the final state $D$ meson is nearly collinear, a hard gluon is required to transfer the large momentum to the spectator charm quark. We choose to employ the PQCD approach based on $K_T$ factorization due to the absence of endpoint singularities. Consequently, Feynman diagrams of factorizable, non-factorizable, and annihilation types can all be computed using this method. The PQCD approach proves effective in calculating non-factorizable and annihilation diagrams and has successfully predicted decays such as $B\to J/\psi D$ \cite{Li:2005vr} and $B^0\to D_{s}^{-}K^{+}$ \cite{Li:2003wg}.
 
The complexity involved in studying the decays of three-body hadronic B mesons is universally recognized as being significantly greater than that of two-body decays. Fortunately, it has been observed that a majority of these decays are predominantly governed by low-energy scalar, vector, and tensor resonance states, which can be effectively described within the framework of quasi-two-body decay \cite{Wang:2016rlo,Ma:2017kec}.
CP violation is a captivating phenomenon in the realm of particle physics. The SM provides a comprehensive framework for elucidating CP violation; however, certain unexplained phenomena persist \cite{Cabibbo:1963yz}. Currently, one avenue of investigating CP violation entails exploring the CKM matrix and hadronic matrix elements, while another involves scrutinizing interference arising from different particle types.
The decay branching ratio also exerts a discernible influence on the exploration of CP asymmetry. In the course of our investigation, we unexpectedly discovered that the employed mixing mechanism significantly impacts the branching fractions of certain decay processes.

In this paper, the branching ratios and CP violations in the decay processes of $B_{c}^{+}\to D_{(s)}^{+}\pi^{+}\pi^{-}$ and $B_{c}^{+}\to D_{(s)}^{+}K^{+}K^{-}$ are investigated using the quasi-two-body method within the framework of PQCD. 
In PQCD, the non-perturbative contributions are exponentially suppressed by Sudakov factors, and the non-perturbative effects are encapsulated within the hadronic wave function \cite{Ali:2007ff}. The hadronic wave function is determined experimentally.
We introduce the mixing mechanism of $\rho^{0}-\omega-\phi$ to compute the branching ratios and CP violations in the quasi-two-body approach. When calculating $V\to \pi^+\pi^-$ and $V\to K^+K^-$, we take into account the mixed resonance effect of $\rho^{0}$, $\omega$, and $\phi$ because of their similar masses.
The positive and negative electrons annihilate into photons and then they are polarized in a vacuum to form the mesons of $\phi (1020)$, $\rho^0(770)$ and $\omega(782)$, which can also decay into $\pi^+\pi^-$ $(K^{+} K^{-})$ pair. Meanwhile, the momentum can also be passed through the VMD model \cite{Ivanov:1981wf,Achasov:2016lbc}. Since the intermediate state particle is an un-physical state, we need convert it into a physical field from an isospin field through the matrix R \cite{Lu:2022rdi}. Then we can obtain the physical state of  $\rho^{0}$, $\omega$ and $\phi$. What deserved to mentioned is that there is no $\rho^{0}-\omega-\phi$ mixing in the physical state and we neglect the contribution of the high-order terms \cite{Shi:2022ggo}.
The physical states $\rho^{0}-\omega-\phi$
can be expressed as linear combinations of the isospin states
$\rho^{0}_{I}-\omega_{I}-\phi_{I}$. The change between the physical field and the isospin field in the intermediate state of the decay process is related by the matrices R.
The off-diagonal elements of R present the information of $\rho^{0}-\omega-\phi$ mixing.
Based on the isospin representation of $\phi_{I}$, $\rho_{I}$ and $\omega_{I}$, the isospin vector $|I,I_{3}>$ can be constructed,
where $I_3$ denotes the third component of isospin.
At present, the Large Hadron Collider (LHC) possesses high energy and high luminosity, and is capable of collecting approximately $10^9$ $B_{c}$ meson events annually \cite{Glasman:2009rqa}. With the upcoming LHC run-3, an increasing amount of $B_{c}$ decay data involving charm resonance states and harmonic states will be gathered to verify the outcomes of our theoretical predictions.     

This paper is structured as follows: In Sec. II, we investigate the CP violation in the decay process $B_c^+ \rightarrow D_{(s)}^+ \pi^+ \pi^-$ via a mixing mechanism involving three vector mesons. In Sec. III, we extend this analysis to the decay process $B_c^+ \rightarrow D_{(s)}^+ K^+ K^-$. In Sec. IV, we derive an integral form representation of local CP violation. In Sec. V, we introduce a formalism for localized CP violation and provide a comparative analysis of the data outcomes. In Sec. VI, we examine the branching ratio of $B_c^+ \rightarrow D_{(s)}^+ K^+ K^-$. Finally, we present a comprehensive summary and conclusion.

\section{CP violation in $\ B_{c}^{+} \rightarrow D_{(s)}^{+}\rho^{0}$ ($\omega$, $\phi$) $   \rightarrow  D_{(s)}^{+}\pi^{+}\pi^{-}$ decay process }
\subsection{The resonance effect from $V\rightarrow \pi^{+}\pi^{-}$ }
\begin{figure}[h]
	\centering
	\includegraphics[height=4cm,width=1.0\textwidth]{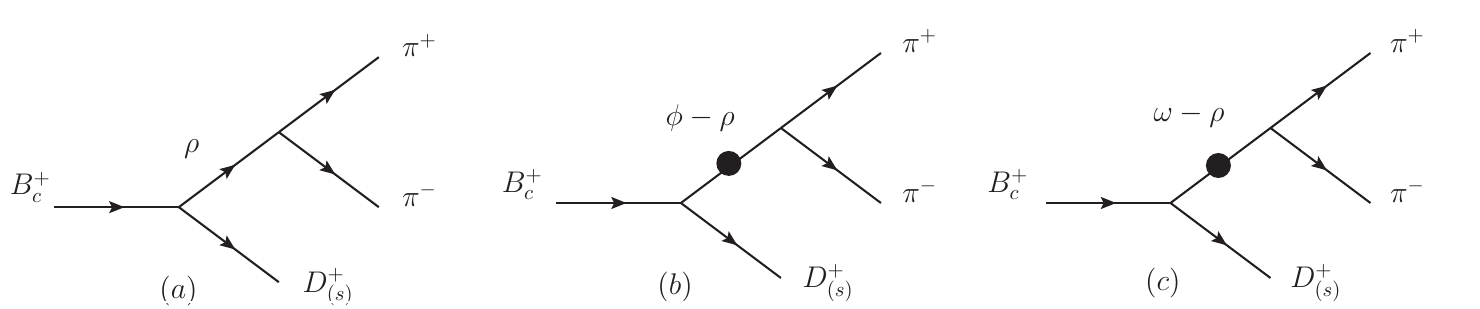}
	\caption{ The decay diagrams of $\ B_{c}^{+} \rightarrow D_{(s)}^{+}\rho^{0}$ ( $\omega$,$\phi$) $   \rightarrow  D_{(s)}^{+}\pi^{+}\pi^{-}$ process.}
	\label{fig1}
\end{figure}
\label{sec:spectra}

We present decay diagrams of the $B_{c}^{+} \rightarrow D_{(s)}^{+}\rho^{0}$ ($\omega$, $\phi$) $\rightarrow D_{(s)}^{+}\pi^{+}\pi^{-}$ process in Fig. 1, aiming to provide a more comprehensive understanding of the mixing mechanism. The quasi-two-body approach used in this study is clearly illustrated in Fig. 1. In these decay diagrams, the processes depicted in (a) represent direct decay modes, where $\pi^{+}\pi^{-}$ pairs are produced via the $\rho^0$ meson. 
Compared to the direct decay processes depicted in diagram (a) of Fig. 1, the $\pi^{+}\pi^{-}$ pair can also be generated through a distinct mixing mechanism. Figs. (b) and (c) indicate that the $\phi$ and $\omega$ mesons undergo resonant decay to a $\pi^{+}\pi^{-}$ meson pair through the mixture with the $\rho^0$ meson. The black dots in the figure represent the resonance effect between these two mesons, denoted by the mixing parameter $\Pi_{V_{i} V_{j}}$.
Although the contribution from this mixing mechanism is relatively small compared to that of diagram (a) in Fig. 1, it must still be considered. Given that the branching ratio of $ \rho^0 \rightarrow \pi^{+} \pi^{-}$ is approximately $100\%$, we can neglect the contributions from $\phi \rightarrow \pi^{+} \pi^{-}$ and $\omega \rightarrow \pi^{+} \pi^{-}$. Consequently, since the direct decays of $\phi \rightarrow \pi^{+} \pi^{-}$ and $\omega \rightarrow \pi^{+} \pi^{-}$ are negligible, the even smaller contribution from resonant mixing can also be disregarded.

The amplitude of the $B_{c}^{+} \rightarrow D_{(s)}^{+}\rho^{0}$ ($\omega$, $\phi$) $\rightarrow D_{(s)}^{+}\pi^{+}\pi^{-}$ decay channel can be described as follows:
\begin{equation}
	A=\left \langle D_{(s)}^{+}\pi^{+}\pi^{-}\left | H^{T} \right | \ B_{c}^{+} \right \rangle +\left \langle D_{(s)}^{+}\pi^{+}\pi^{-}\left | H^{P} \right |  B_{c}^{+} \right \rangle,
\end{equation}
where $\left \langle D_{(s)}^{+}\pi^{+}\pi^{-}\left | H^{P} \right | \ B_{c}^{+} \right \rangle $ and $\left \langle D_{(s)}^{+}\pi^{+}\pi^{-}\left | H^{T} \right | \ B_{c}^{+} \right \rangle $ represent the amplitudes associated with penguin-level and tree-level contributions, respectively.
Neglecting higher order terms, the amplitudes
can be as demonstrated below:
\begin{eqnarray}
	\begin{split}
		\left \langle D_{(s)}^{+}\pi^{+}\pi^{-}\left | H^{T} \right | \ B_{c}^{+} \right \rangle=
		&
	    \frac{g_{\rho^0 \rightarrow \pi^{+} \pi^{-}}}{({s-m_{\rho ^0}^{2}+im_{\rho ^0} \varGamma _{\rho ^0}})}t_{\rho}
		+\frac{g_{\rho^0 \rightarrow \pi^{+} \pi^{-}}}{({s-m_{\rho ^0}^{2}+im_{\rho ^0} \varGamma _{\rho ^0}})({s-m_{\omega}^{2}+im_{\omega} \varGamma_{\omega}})}\widetilde{\Pi}_{\rho\omega}t_{\omega} \\&
		+\frac{g_{\rho^0 \rightarrow \pi^{+} \pi^{-}}}{({s-m_{\rho ^0}^{2}+im_{\rho ^0} \varGamma _{\rho ^0}})({s-m_{\phi}^{2}+im_{\phi} \varGamma_{\phi}})}\widetilde{\Pi}_{\rho\phi}t_{\phi},
		\label{Htr}
	\end{split}
\end{eqnarray}
\begin{eqnarray}
	\begin{split}
		\left \langle D_{(s)}^{+}\pi^{+}\pi^{-}\left | H^{P} \right | \ B_{c}^{+} \right \rangle=
		&
		\frac{g_{\rho^0 \rightarrow \pi^{+} \pi^{-}}}{({s-m_{\rho ^0}^{2}+im_{\rho ^0} \varGamma _{\rho ^0}})}p_{\rho}
		+\frac{g_{\rho^0 \rightarrow \pi^{+} \pi^{-}}}{({s-m_{\rho ^0}^{2}+im_{\rho ^0} \varGamma _{\rho ^0}})({s-m_{\omega}^{2}+im_{\omega} \varGamma_{\omega}})}\widetilde{\Pi}_{\rho\omega}p_{\omega} \\&
		+\frac{g_{\rho^0 \rightarrow \pi^{+} \pi^{-}}}{({s-m_{\rho ^0}^{2}+im_{\rho ^0} \varGamma _{\rho ^0}})({s-m_{\phi}^{2}+im_{\phi} \varGamma_{\phi}})}\widetilde{\Pi}_{\rho\phi}p_{\phi},
		\label{Hpe}
	\end{split}
\end{eqnarray}
where the tree-level (penguin-level) amplitudes $t_{\rho}\left(p_{\rho}\right)$, $t_{\omega}\left(p_{\omega}\right)$, and $t_{\phi}\left(p_{\phi}\right)$ correspond to the decay processes $\ B_{c}^{+} \rightarrow D_{(s)}^{+}\rho^0 $, $\ B_{c}^{+} \rightarrow D_{(s)}^{+}\omega $ and $\ B_{c}^{+} \rightarrow D_{(s)}^{+}\phi $, respectively.  
 Here, $s_V$ denotes the inverse propagator of the vector meson V, defined as $s_{V}=s-m_{V}^{2}+\mathrm{i} m_{V} \Gamma_{V}$ \cite{Chen:1999nxa}. The parameters $m_V$ and $\Gamma_{V}$ correspond to the mass and decay width of the vector mesons, respectively. Additionally, $\sqrt{s}$ represents the invariant mass of the $\pi^+\pi^-$ system. 
Moreover, $g_{\rho\rightarrow \pi^{+} \pi^{-}}$ represents the coupling constant derived from the decay process of $ \rho^0 \rightarrow \pi^{+} \pi^{-}$.
In this paper, the momentum dependence of the mixing parameters $\Pi_{V_{i} V_{j}}$ of $V_{i}V_{j}$ mixing is introduced to obtain the obvious s dependence. 
The mixing parameter $\Pi_{\rho \omega }=(-4470 \pm 250 \pm 160) - i(5800 \pm 2000 \pm 1100) \, \mathrm{MeV}^{2}$ is recently determined with high precision near the $\rho$ meson by Wolfe and Maltnan \cite{Lu:2018fqe,Wolfe:2009ts,Wolfe:2010gf}. The mixing parameter $\Pi_{\omega \phi}=(19000 + i(2500 \pm 300)) \, \mathrm{MeV}^{2}$ is obtained near the $\phi$ meson. Additionally, the mixing parameter $\Pi_{\phi\rho}=(720 \pm 180) - i(870 \pm 320) \, \mathrm{MeV}^{2}$ is measured near the $\phi$ meson \cite{Achasov:1999wr}. We then define the following renormalized mixing parameters:
\begin{eqnarray}
	\widetilde{\Pi}_{\rho\omega}=\frac{s_{\rho}\Pi_{\rho\omega}}{s_{\rho}-s_{\omega}},~~
	\widetilde{\Pi}_{\rho\phi}=\frac{s_{\rho}\Pi_{\rho\phi}}{s_{\rho}-s_{\phi}},~~
	\widetilde{\Pi}_{\phi\omega}=\frac{s_{\phi}\Pi_{\phi\omega}}{s_{\phi}-s_{\omega}}.
\end{eqnarray}

The differential parameter for CP asymmetry can be expressed as follows:
\begin{equation}
	\label{cp31}
	A_{CP}=\frac{\left| A \right|^2-\left| \overline{A} \right|^2}{\left| A \right|^2+\left| \overline{A} \right|^2}.
\end{equation}

\subsection{Formulation of calculations}

The three-body decay process involves complex and multifaceted dynamical mechanisms. The PQCD method is renowned for its effectiveness in addressing perturbative corrections, a capability that has been successfully demonstrated in two-body non-leptonic decays and shows potential for quasi-two-body decays as well.
In the framework of PQCD, within the rest frame of a heavy B meson, the decay process involves the production of two light mesons with significantly large momenta, indicating rapid motion. The dominance of hard interactions in this decay amplitude is attributed to the insufficient time for soft gluon exchanges with the final-state mesons. Given the high velocities of these final-state mesons, a hard gluon imparts momentum to the light spectator quark within the B meson, leading to the formation of rapidly moving final-state mesons. Consequently, this hard interaction is characterized by six-quark operators. The non-perturbative dynamics are encapsulated within the meson wave function, which can be determined through experimental measurements. On the other hand, perturbation theory enables the computation of the aforementioned hard contribution. Quasi-two-body decay can be analyzed by defining the intermediate state of the decay process.
In accordance with the concept of quasi-two-body decay, the Feynman diagram representing the amplitude for a three-body decay can be constructed by applying the Feynman rules.

\begin{figure}[h]
	\centering
	\includegraphics[height=6cm,width=1.0\textwidth]{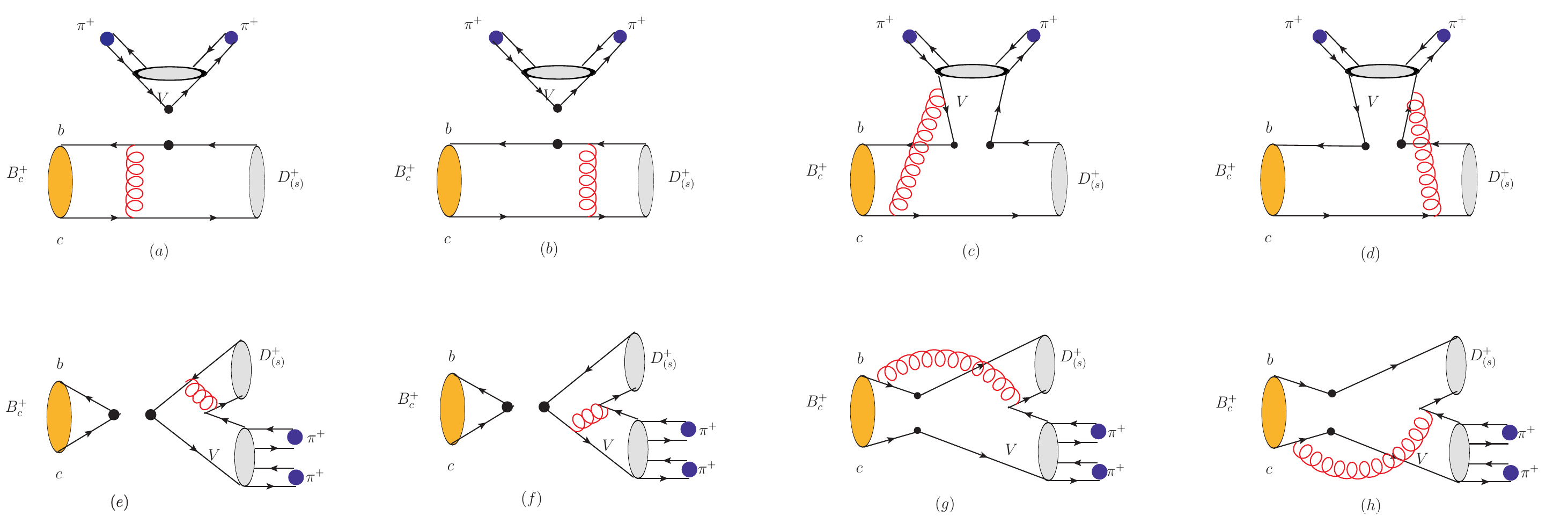}
	\caption{ The Feynman diagrams for emission and annihilation contributions in the $\ B_{c}^{+} \rightarrow D_{(s)}^{+}V\rightarrow  D_{(s)}^{+}\pi^{+}\pi^{-}$ decay process.}
	\label{fig1}
\end{figure}

In Fig. 2, diagrams (a) and (b) illustrate the contributions from factorizable emission processes in the $B_{c}^{+}$ meson decay, while diagrams (c) and (d) depict the contributions from non-factorizable emission processes. Diagrams (e) and (f) show the contributions from factorizable annihilation processes, whereas diagrams (g) and (h) highlight the contributions from non-factorizable annihilation processes.

By employing the quasi-two-body decay method, the total amplitude of  $\ B_{c}^{+} \rightarrow D_{(s)}^{+}\rho^{0}$ ($\omega$, $\phi$) $ $ $\rightarrow D_{(s)}^{+}\pi^{+}\pi^{-}$ is composed of two components:$\ B_{c}^{+} \rightarrow D_{(s)}^{+} \rho^{0}$ ($\omega$, $\phi$) and $\rho^0 $ ($\omega$, $\phi) \rightarrow \pi^{+}\pi^{-}$. In this study, we illustrate the methodology of quasi-two-body decay process using the example of $\ B_{c}^{+} \rightarrow D^{+} \rho^{0}$ $\rightarrow D^{+}\pi^{+}\pi^{-}  $, based on the matrix elements involving  $V_{ud}$, $V_{ub}^{*}$ and  $V_{td}$,$V_{tb}^{*}$.
The amplitude of Fig 1(a) is presented as follows:  
\begin{equation}
	\begin{aligned}
		\sqrt{2}A\left(\ B_{c}^{+} \rightarrow\right.\left. D^{+}\rho^{0}\rightarrow D^{+}\pi^{+}\pi^{-}\right) =&  \frac{<D^{+}\rho ^0|H_eff|B_{c}^{+}><\pi^{+}\pi^{-}|H_{\rho ^0\rightarrow \pi ^+\pi ^- }|\rho ^0>}{{s-m_{\rho ^0}^{2}+im_{\rho ^0} \varGamma _{\rho ^0}}}\\ & 
		=\sum_{\lambda =0,\pm 1} \frac{G_{F}P_{(B_{c}^{+}}\cdot \epsilon^*\left( \lambda \right)\ g^{\rho ^0\rightarrow \pi ^+\pi ^- }\epsilon \left( \lambda \right) \cdot \left( p_{\pi ^+}-p_{\pi ^-} \right)}{\sqrt{2}{({s-m_{\rho ^0}^{2}+im_{\rho ^0} \varGamma _{\rho ^0}}})}
		\\& 
		\left.\times \Bigg\{V_{u d} V_{u b}^{*}\left[F_{e}^{L L}(C_1+\frac{1}{3}C_2)+M_{e}^{L L}(C_{2})+F_{a}^{L L }(C_{2}+\frac{1}{3}C_1)+M_{a}^{L L}(C_{1})\right]  \right.\\
		&+ V_{t d} V_{t b}^{*}\bigg\{F_{a}^{L L}(C_{2}+\frac{1}{3}C_1)+M_{a}^{L L}(C_{1})-\left[\mathcal {M}_{e}^{LL}(\frac{3}{2}C_{10}-C_3+\frac{1}{2}C_9)\right.\\
		&\left.-M_{a}^{L L}(C_{3}+C_{9})+M_{e}^{L R}(-C_{5}+\frac{1}{2}C_{7})+(-C_4-\frac{1}{3}C_3-C_{10}-\frac{1}{3}C_9){F}_a^{LL}\right.\\
		&\left.+(C_{10}+\frac{5}{3}C_9-\frac{1}{3}C_3-C_4-\frac{3}{2}C_7-\frac{1}{2}C_8){F}_e^{LL}\right.\\
		&\left.+(-C_6-\frac{1}{3}C_5+\frac{1}{2}C_{8}+\frac{1}{6}C_7){F}_e^{SP}-(C_5+C_7)\mathcal {M}_a^{LR}\right.\\
		&\left.+(-C_6-\frac{1}{3}C_5-C_{8}-\frac{1}{3}C_7) {F}_a^{SP}\right]\bigg\}\Bigg\} ,
	\end{aligned}
\end{equation}
where $P_{B_{c}^{+}}$, $p_{\pi^{+}}$, and $p_{\pi^{-}}$ represent the momenta of $B_{c}^{+}$, $\pi^{+}$, and $\pi^{-}$, respectively. $C_i$ ($a_i$) denotes the Wilson coefficient (associated Wilson coefficient), $\epsilon$ represents the polarization of the vector meson, and $G_F$ is the Fermi constant. $f_{\pi}$ refers to the decay constant of the pion \cite{Li:2006jv}. In the equation, $F_{e}^{LL}$, $F_{e}^{LR}$, and $F_{e}^{SP}$ denote the contributions from factorizable emission diagrams, while $M_{e}^{LL}$, $M_{e}^{LR}$, and $M_{e}^{SP}$ indicate the contributions from non-factorizable emission diagrams. Similarly, $F_{a}^{LL}$, $F_{a}^{LR}$, and $F_{a}^{SP}$, as well as $M_{a}^{LL}$, $M_{a}^{LR}$, and $M_{a}^{SP}$, correspond to the contributions from factorizable and non-factorizable annihilation diagrams, respectively. The terms $LL$, $LR$, and $SP$ refer to three different flow structures.

The additional representation of the three-body decay amplitude that needs to be taken into account when calculating CP violation through the mixing mechanism corresponding to Figs. 1(b) and (c) is presented as follows:
\begin{equation}
	\begin{aligned}
		A\left(\ B_{c}^{+} \rightarrow\right.\left. D^{+}(\phi-\rho^0)\rightarrow D^{+}\pi^{+}\pi^{-}\right)=&\frac{<D^{+}\phi|H_eff|B_{c}^{+}><\pi^{+}\pi^{-}|H_{\rho^{0} \rightarrow \pi^{+} \pi^{-} }|\rho^{0}>\widetilde{\Pi}_{\rho\phi}}{({s-m_{\phi}^{2}+im_{\phi} \varGamma_{\phi}})({s-m_{\rho ^0}^{2}+im_{\rho ^0} \varGamma _{\rho ^0}})} \\&
		=\sum_{\lambda =0,\pm 1} \frac{G_{F}P_{(B_{c}^{+}}\cdot \epsilon^*\left( \lambda \right)\ g^{\rho ^0\rightarrow \pi ^+\pi ^- }\epsilon \left( \lambda \right) \cdot \left( p_{\pi ^+}-p_{\pi ^-} \right)\widetilde{\Pi}_{\rho\phi}}{\sqrt{2}({s-m_{\phi}^{2}+im_{\phi} \varGamma_{\phi}})({s-m_{\rho ^0}^{2}+im_{\rho ^0} \varGamma _{\rho ^0}})}
		\\	&
		\times \bigg\{-V_{u d} V_{u b}^{*}\left[(C_4-\frac{1}{2}C_{10})\mathcal {M}_{e}^{LL}+(C_6-\frac{1}{2}C_8)\mathcal {M}_{e}^{SP}+(C_3+\frac{1}{3}C_4\right.\\    &\left.-\frac{1}{2}C_9-\frac{1}{6}C_{10}){F}_{e}^{LL}+(C_5+\frac{1}{3}C_6-\frac{1}{2}C_7-\frac{1}{6}C_8){F}_{e}^{LR}\right]\bigg\},
	\end{aligned}
\end{equation}
and
\begin{equation}
\begin{aligned}
\sqrt{2}A\left(\ B_{c}^{+} \rightarrow\right.\left. D^{+}(\omega-\rho^0)\rightarrow D^{+}\pi^{+}\pi^{-}\right)=&\frac{<D^{+}\omega|H_eff|B_{c}^{+}><\pi^{+}\pi^{-}|H_{\rho^{0} \rightarrow \pi^{+} \pi^{-} }|\rho^{0}>\widetilde{\Pi}_{\rho\omega}}{({s-m_{\omega}^{2}+im_{\omega} \varGamma_{\omega}})({s-m_{\rho ^0}^{2}+im_{\rho ^0} \varGamma _{\rho ^0}})} \\&
=\sum_{\lambda =0,\pm 1} \frac{G_{F}P_{(B_{c}^{+}}\cdot \epsilon^*\left( \lambda \right)\ g^{\rho ^0\rightarrow \pi ^+\pi ^- }\epsilon \left( \lambda \right) \cdot \left( p_{\pi ^+}-p_{\pi ^-} \right)\widetilde{\Pi}_{\rho\omega}}{\sqrt{2}({s-m_{\omega}^{2}+im_{\omega} \varGamma_{\omega}})({s-m_{\rho ^0}^{2}+im_{\rho ^0} \varGamma _{\rho ^0}})}\\	&
\left.\times \bigg\{V_{u d} V_{u b}^{*}\left[ F_{e}^{L L}\left(c_{1}+\frac{1}{3}C_2\right)+C_2\mathcal {M}_{e}^{LL}-\left((C_2+\frac{1}{3}C_1)F_{a}^{L L}+C_1\mathcal {M}_{a}^{LL}\right)\right]\right.\\
&\left.-V_{t d} V_{t b}^{*}\left[(C_2+\frac{1}{3}C_1)F_{a}^{L L}+C_1\mathcal {M}_{a}^{LL}+\left((2C_4+C_3+\frac{1}{2}C_{10}-\frac{1}{2}C_9)\mathcal {M}_{e}^{LL}\right)\right.\right.\\
&\left.+(C_3+C_9)\mathcal {M}_{e}^{LL}+(C_5-\frac{1}{2}C_7)\mathcal {M}_{e}^{LR}+(C_5+C_7)\mathcal {M}_{a}^{LR}\right.\\
&\left.+(C_4+\frac{1}{3}C_3+C_{10}+\frac{1}{3}C_9) {F}_{a}^{LL}+(\frac{7}{3}C_3+\frac{5}{3}C_4+\frac{1}{3}(C_9-C_{10})){F}_{e}^{LL}\right.\\
&\left.+(2C_5+\frac{2}{3}C_6+\frac{1}{2}C_7+\frac{1}{6}C_8){F}_{e}^{LR}+(C_6+\frac{1}{3}C_5-\frac{1}{2}C_{8}-\frac{1}{6}C_7){F}_{e}^{SP}\right.\\
&\left.+(C_6+\frac{1}{3}C_5+C_{8}+\frac{1}{3}C_7){F}_{a}^{SP}\right]\bigg\}.
\end{aligned}
\end{equation}

The following equations represent the amplitude forms of the three-body decay $B_{c}^{+} \rightarrow D_{s}^{+}\pi^{+}\pi^{-}$, as illustrated in Fig. 1:
\begin{equation}
	\begin{aligned}
		\sqrt{2}A\left(\ B_{c}^{+} \rightarrow\right.\left. D_{s}^{+}\rho^{0}\rightarrow D_{s}^{+}\pi^{+}\pi^{-}\right) =&  \frac{<D_{s}^{+}\rho ^0|H_eff|B_{c}^{+}><\pi^{+}\pi^{-}|H_{\rho ^0\rightarrow \pi ^+\pi ^- }|\rho ^0>}{{s-m_{\rho ^0}^{2}+im_{\rho ^0} \varGamma _{\rho ^0}}}\\ & 
		=\sum_{\lambda =0,\pm 1} \frac{G_{F}P_{(B_{c}^{+}}\cdot \epsilon^*\left( \lambda \right)\ g^{\rho ^0\rightarrow \pi ^+\pi ^- }\epsilon \left( \lambda \right) \cdot \left( p_{\pi ^+}-p_{\pi ^-} \right)}{\sqrt{2}{({s-m_{\rho ^0}^{2}+im_{\rho ^0} \varGamma _{\rho ^0}}})}
		\\&
		 \times \left\{  V_{u s} V_{u b}^{*}\left[ F_e^{L L}\left(c_{1}+\frac{1}{3}C_{2}\right)+\mathcal{M}_e^{L L}\left(C_{2}\right)\right]- V_{t s} V_{t b}^{*}\left[ \left(\frac{1}{2}(3C_9+C_{10}){F}_e^{LL}\right)\right.\right.\\
		&\left.\left.\frac{1}{2}(3C_7+C_8){F}_e^{LR}+\frac{3}{2}C_{10}\mathcal {M}_e^{LL}+\frac{3}{2}C_{8}\mathcal {M}_e^{SP}\right]\right\},
	\end{aligned}
\end{equation}

\begin{equation}
\begin{aligned}
A\left(\ B_{c}^{+} \rightarrow\right.\left. D_{s}^{+}(\phi-\rho^0)\rightarrow D_{s}^{+}\pi^{+}\pi^{-}\right)=&\frac{<D_{s}^{+}\phi|H_eff|B_{c}^{+}><\pi^{+}\pi^{-}|H_{\rho^{0} \rightarrow \pi^{+} \pi^{-} }|\rho^{0}>\widetilde{\Pi}_{\rho\phi}}{({s-m_{\phi}^{2}+im_{\phi} \varGamma_{\phi}})({s-m_{\rho ^0}^{2}+im_{\rho ^0} \varGamma _{\rho ^0}})} \\&
=\sum_{\lambda =0,\pm 1} \frac{G_{F}P_{(B_{c}^{+}}\cdot \epsilon^*\left( \lambda \right)\ g^{\rho ^0\rightarrow \pi ^+\pi ^- }\epsilon \left( \lambda \right) \cdot \left( p_{\pi ^+}-p_{\pi ^-} \right)\widetilde{\Pi}_{\rho\phi}}{\sqrt{2}({s-m_{\phi}^{2}+im_{\phi} \varGamma_{\phi}})({s-m_{\rho ^0}^{2}+im_{\rho ^0} \varGamma _{\rho ^0}})}
\\	&
\times \left\{- V_{u s} V_{u b}^{*}\left[F_a^{L L}\left(c_{2}+\frac{1}{3} c_{1}\right)+\mathcal{M}_a^{L L}c_{1}\right]  \right.\\
&\left. -V_{t s} V_{t b}^{*}\left[F_a^{L L}\left(c_{2}+\frac{1}{3} c_{1}\right)+\mathcal{M}_a^{L L}c_{1}+M_e^{L L}\left(C_{3}+C_{4}-\left(\frac{1}{2} C_{9}+C_{10}\right)\right)\right.\right.\\
&\left.+M_a^{L L}\left(C_{3}+C_{9}\right)+M_e^{L R}\left(C_{5}-\frac{1}{2} C_{7}\right)+M_a^{L R}\left(C_{5}+C_{7}\right) \right.\\
&\left. F_a^{L L}\left(C_{4}+\frac{1}{3} C_{3}+ C_{10}+\frac{1}{3} C_{9}\right)+\mathcal{M}_e^{S P}\left(C_{6}-\frac{1}{2} C_{8}\right)+F_{e}^{L L}\frac{2}{3}\left(2\left(C_3+C_4\right)\right.\right.\\
&\left.\left.-\left(C_9+C_{10}\right)\right)+(C_5+\frac{1}{3}C_6-\frac{1}{2}C_7-\frac{1}{6}C_8){F}_{e}^{LR}+\right.\\
&\left.\left. (C_6+\frac{1}{3}C_5-\frac{1}{2}C_{8}-\frac{1}{6}C_7){F}_{e}^{SP}+(C_6+\frac{1}{3}C_5+C_{8}+\frac{1}{3}C_7){F}_{a}^{SP} \right]\right\},
\end{aligned}
\end{equation}
and
\begin{equation}
	\begin{aligned}
		\sqrt{2}A\left(\ B_{c}^{+} \rightarrow\right.\left. D_{s}^{+}(\omega-\rho^0)\rightarrow D_{s}^{+}\pi^{+}\pi^{-}\right)=&\frac{<D_{s}^{+}\omega|H_eff|B_{c}^{+}><\pi^{+}\pi^{-}|H_{\rho^{0} \rightarrow \pi^{+} \pi^{-} }|\rho^{0}>\widetilde{\Pi}_{\rho\omega}}{({s-m_{\omega}^{2}+im_{\omega} \varGamma_{\omega}})({s-m_{\rho ^0}^{2}+im_{\rho ^0} \varGamma _{\rho ^0}})} \\&
		=\sum_{\lambda =0,\pm 1} \frac{G_{F}P_{(B_{c}^{+}}\cdot \epsilon^*\left( \lambda \right)\ g^{\rho ^0\rightarrow \pi ^+\pi ^- }\epsilon \left( \lambda \right) \cdot \left( p_{\pi ^+}-p_{\pi ^-} \right)\widetilde{\Pi}_{\rho\omega}}{\sqrt{2}({s-m_{\omega}^{2}+im_{\omega} \varGamma_{\omega}})({s-m_{\rho ^0}^{2}+im_{\rho ^0} \varGamma _{\rho ^0}})}\\	&
		\times  \left\{V_{u s} V_{u b}^{*}  \left[ F_e^{L L}\left(c_{1}+\frac{1}{3}C_{2}\right)+\mathcal{M}_e^{L L}\left(C_{2}\right)\right] - V_{t s} V_{t b}^{*}\left[(2C_4+\frac{1}{2}C_{10})\mathcal {M}_{e}^{LL} \right.\right.\\
		&\left.(2C_6+\frac{1}{2}C_8)\mathcal {M}_{e}^{SP}+
		(2C_3+\frac{2}{3}C_4+\frac{1}{2}C_9+\frac{1}{6}C_{10}){F}_{e}^{LL}\right.\\
		&\left.\left.+(2C_5+\frac{2}{3}C_6+\frac{1}{2}C_7+\frac{1}{6}C_8){F}_{e}^{LR}\right]\right\}.
	\end{aligned}
\end{equation}

Hence, the total amplitude for the process $B_{c}^{+} \rightarrow D_{(s)}^{+}\rho^{0} (\omega, \phi) \rightarrow D_{(s)}^{+}\pi^{+}\pi^{-}$ is the coherent sum of the amplitudes from the three diagrams (a), (b), and (c) shown in Fig. 1. This includes both the direct decay amplitudes and all contributions from mixed resonance states.

\subsection{CP violation results of the $\ B_{c}^{+} \rightarrow   D_{(s)}^{+}\pi^{+}\pi^-$  decay process}

\begin{figure}[h]
	\centering
	\begin{minipage}[h]{0.45\textwidth}
		\centering
		\includegraphics[height=4cm,width=6.5cm]{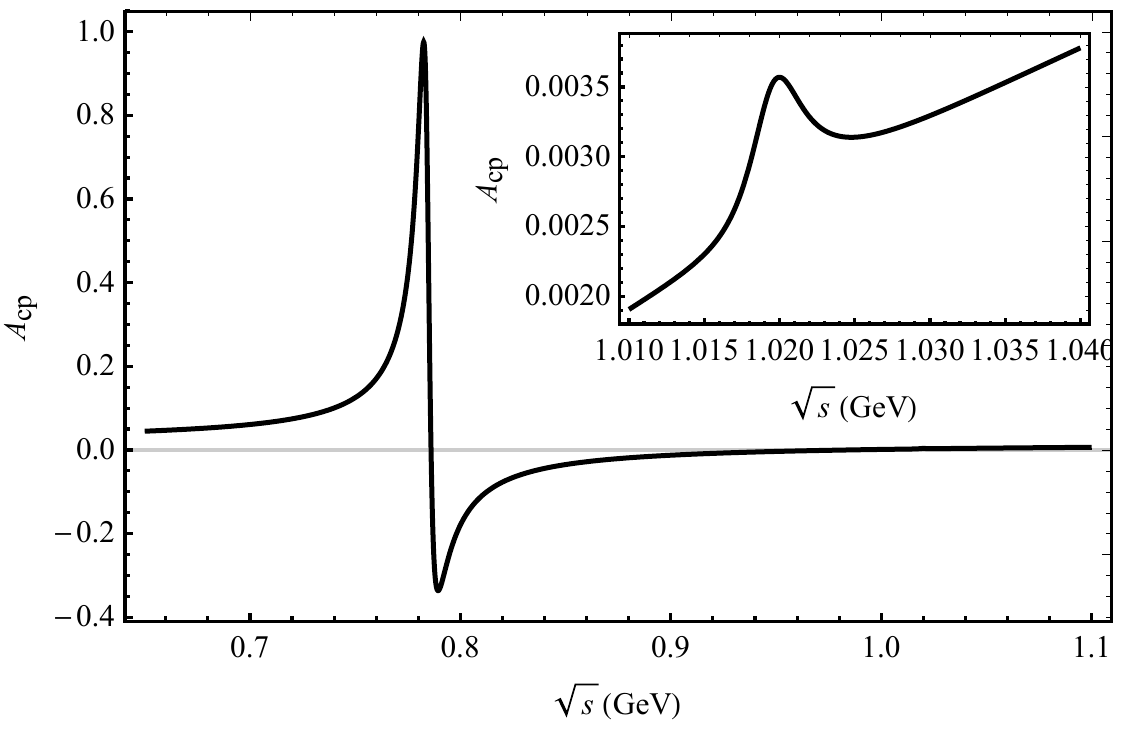}
		\caption{Plot of  $A_{CP}$ as a function of $\sqrt{s}$ corresponding to central parameter values of CKM matrix elements
			for the decay channel of $ \ B_{c}^{+} \rightarrow D^{+}\pi^{+}\pi^{-}$.}
		\label{fig2}
	\end{minipage}
	\quad
	\begin{minipage}[h]{0.45\textwidth}
		\centering
		\includegraphics[height=4cm,width=6.5cm]{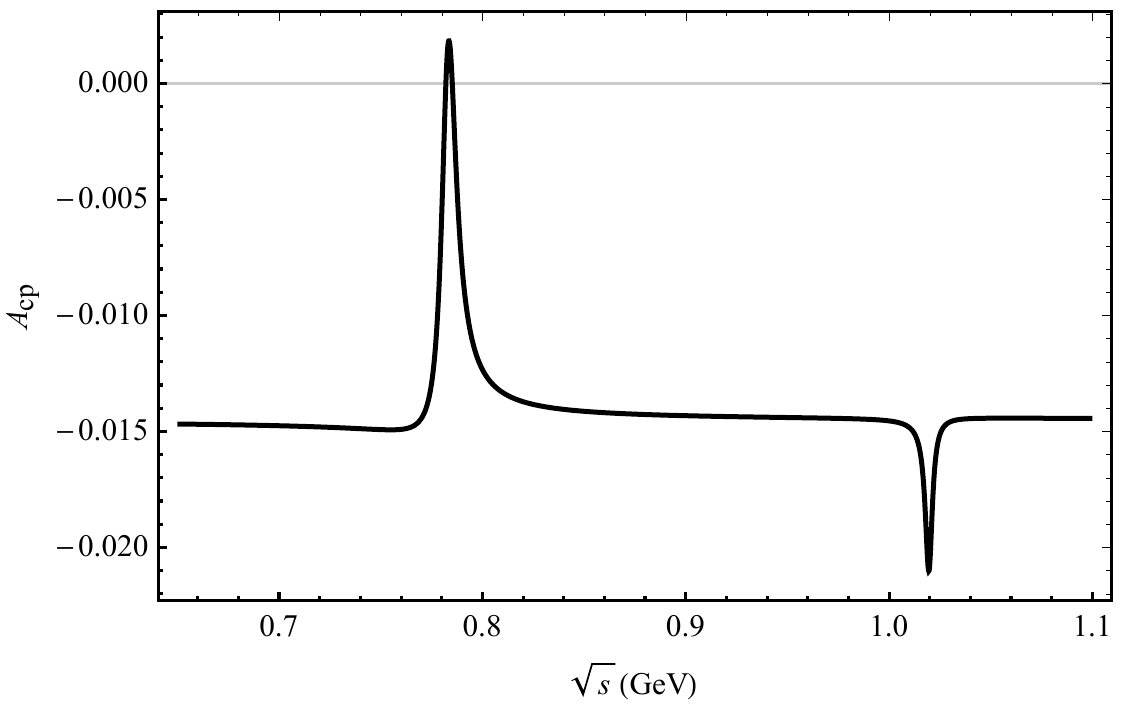}
		\caption{Plot of  $A_{CP}$ as a function of $\sqrt{s}$ corresponding to central parameter values of CKM matrix elements
			for the decay channel of $ B_{c}^{+} \rightarrow D_s^{+}\pi^{+}\pi^{-}$.}
		\label{fig3}
	\end{minipage}
\end{figure}

We present the outcome plots that illustrate CP violation in the decay processes of $B_{c}^{+} \rightarrow D^{+}\pi^{+}\pi^{-}$ and $B_{c}^{+} \rightarrow D_s^{+}\pi^{+}\pi^{-}$.
We present outcome plots that illustrate CP violation in the decay processes of $B_{c}^{+} \rightarrow D^{+}\pi^{+}\pi^{-}$ and $B_{c}^{+} \rightarrow D_s^{+}\pi^{+}\pi^{-}$. As shown in the figures, we investigate the mixing of $\rho-\omega-\phi$ particles. Figs. 3 and 4 depict the variation of $A_{CP}$ as a function of $\sqrt{s}$, which represents the invariant mass of the $\pi^{+}\pi^{-}$ system. The central parameter values of the CKM matrix elements are utilized to derive these results. The observed CP violation in these decay processes offers valuable insights into fundamental physics phenomena, including vector meson interferences.

The maximum of CP violation from the decay process $ \ B_{c}^{+} \rightarrow D^{+}\pi^{+}\pi^{-}$ in Fig. 3, with a value of $97.54\%$, occurs at an invariant mass of 0.782 GeV, which corresponds to the mass position of the $\rho$ and $\omega$ meson. Additionally, small peaks are also observed in the invariant mass range of $\phi$(1.02 Gev). The value of CP violation is rather small, with a magnitude of $0.357\%$. In the decay process of $ \ B_{c}^{+} \rightarrow D^{+}\phi$, only the contribution from the penguin diagram exists, while there is no contribution from the tree diagram. Hence, it can be demonstrated that the mixed resonance of $\rho^0 -\omega$ makes a considerable contribution in the decay process. 

Regarding the decay process of $B_{c}^{+} \rightarrow D_s^{+}\pi^{+}\pi^{-}$ illustrated in Fig. 4, we have observed a trend similar to that of $B_{c}^{+} \rightarrow D^{+}\pi^{+}\pi^{-}$. When the invariant mass of $\pi^{+}\pi^{-}$ approaches that of the $\rho^0$ or $\omega$, a pronounced peak is observed at approximately 0.783 GeV with a CP violation of $0.18\%$. Additionally, there is a smaller peak near the mass of the $\phi$ meson, corresponding to a value of $-2.10\%$.

\section{CP violation in $\ B_{c}^{+} \rightarrow D_{(s)}^{+}\phi$ ($\rho^0$, $\omega$) $   \rightarrow  D_{(s)}^{+}K^{+}K^{-}$ decay process }
\subsection{The resonance effect from $V\rightarrow K^{+}K^{-}$}
\begin{figure}[h]
	\centering
	\includegraphics[height=11cm,width=1.0\textwidth]{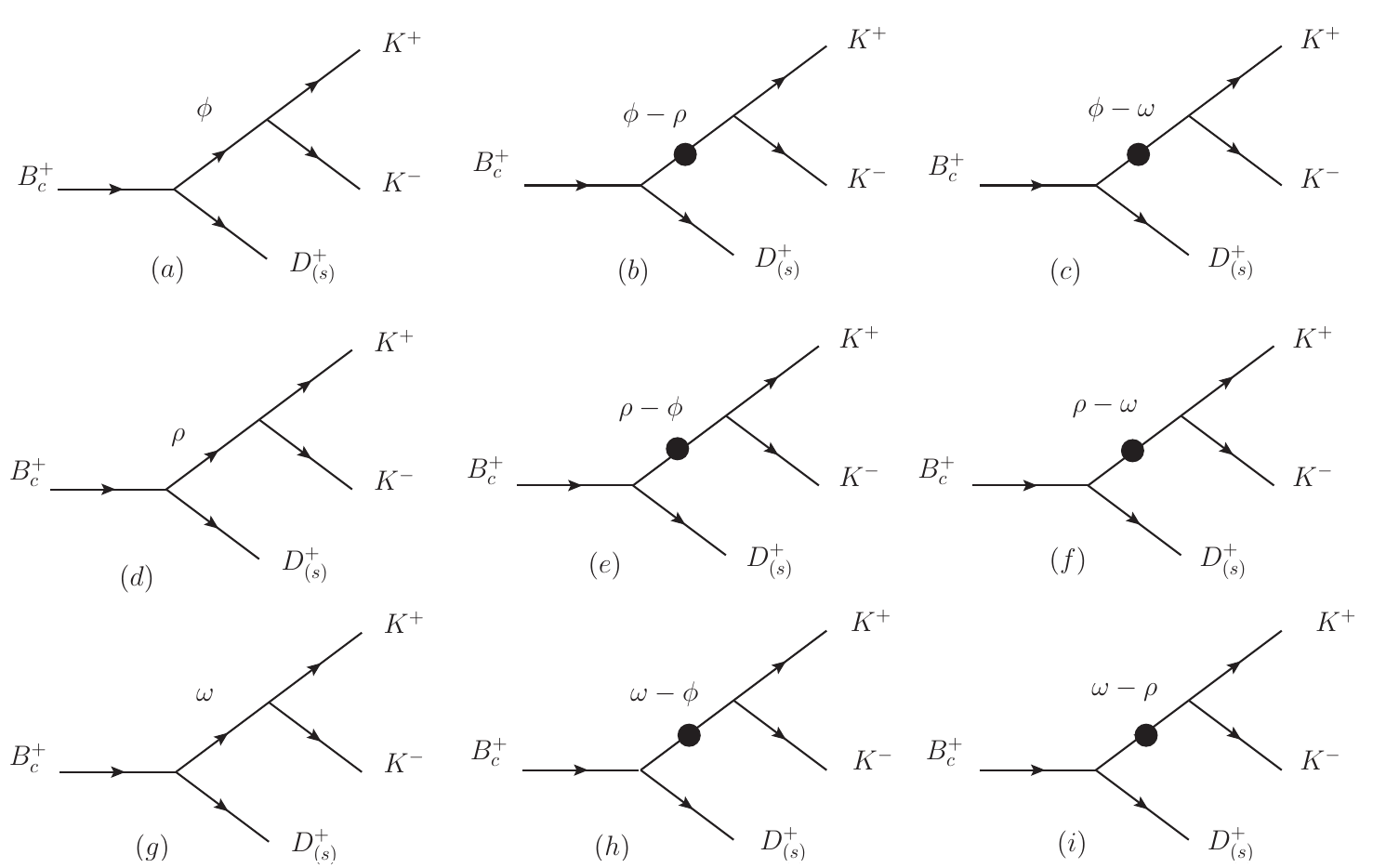}
	\caption{ The decay diagrams of $\ B_{c}^{+} \rightarrow D_{(s)}^{+}\phi$ ($\rho^0$, $\omega$)  $   \rightarrow  D_{(s)}^{+}K^{+}K^{-}$ process.}
	\label{fig1}
\end{figure}
\label{sec:spectra}
We present the decay diagrams of the $B_{c}^{+} \rightarrow D_{(s)}^{+}\phi$ ($\rho^0$, $\omega$) $\rightarrow D_{(s)}^{+}K^{+}K^{-}$ process in Fig. 5. In these decay diagrams, the processes depicted in (a), (d), and (g) represent direct decay modes, where the $K^{+} K^{-}$ pairs are produced via the intermediate states of $\rho^0$, $\omega$, and $\phi$, respectively. Unlike the decay to $\pi^{+} \pi^{-}$ meson pairs, all three vector mesons $\rho^0$, $\omega$, and $\phi$ can directly decay into $K^{+} K^{-}$ meson pairs. In addition to the direct decay mechanism, the $K^{+} K^{-}$ meson pair can also be generated through a mixing mechanism. 
Fig. 5(b), (c), (e), (f), (h), and (i) illustrate the contributions of the $(\phi-\rho^0)$,$(\phi-\omega)$, $(\rho^0-\phi)$, $(\rho^0-\omega)$, $(\omega-\phi)$, and $(\omega-\rho^0)$ resonance hybrids, respectively. For instance, Fig. 5(b) demonstrates that the $B_{c}^{+}$ meson decays into the $D_{(s)}^{+}$ and $\phi$ mesons via quasi-two-body decay, followed by the resonant decay of the $\phi$ meson into a $K^{+} K^{-}$ meson pair through mixing with the $\rho^0$ meson. Consequently, the $K^{+} K^{-}$ meson pairs generated by this mixing mechanism correspond to six distinct cases. Although the contribution from this mixing mechanism is relatively small compared to direct decay, it remains significant and must be considered. Similar to the decay channel of $B_{c}^{+} \rightarrow D_{(s)}^{+}\pi^{+}\pi^-$, we provide the amplitude for $B_{c}^{+} \rightarrow D_{(s)}^{+}K^{+}K^-$:
\begin{eqnarray}
	\begin{split}
		\left \langle D_{(s)}^{+}K^{+}K^{-}\left | H^{T} \right | \ B_{c}^{+} \right \rangle=
		&
		\frac{g_{\phi\rightarrow K^{+} K^{-}}}{s_{\phi}}t_{\phi}
		+\frac{g_{\rho\rightarrow K^{+} K^{-}}}{s_{\rho}s_{\phi}}\widetilde{\Pi}_{\rho\phi}t_{\phi}
		+\frac{g_{\omega\rightarrow K^{+} K^{-}}}{s_{\omega}s_{\phi}}\widetilde{\Pi}_{\omega\phi}t_{\phi}
		+\frac{g_{\rho\rightarrow K^{+} K^{-}}}{s_{\rho}}t_{\rho}
		+\frac{g_{\phi\rightarrow K^{+} K^{-}}}{s_{\phi}s_{\rho}}\widetilde{\Pi}_{\phi\rho}t_{\rho}
		\\
		&
		+\frac{g_{\omega\rightarrow K^{+} K^{-}}}{s_{\omega}s_{\rho}}\widetilde{\Pi}_{\omega\rho}t_{\rho}
		+\frac{g_{\omega\rightarrow K^{+} K^{-}}}{s_{\omega}}t_{\omega}
		+\frac{g_{\phi\rightarrow K^{+} K^{-}}}{s_{\phi}s_{\omega}}\widetilde{\Pi}_{\phi\omega}t_{\omega}
		+\frac{g_{\rho\rightarrow K^{+} K^{-}}}{s_{\rho}s_{\omega}}\widetilde{\Pi}_{\rho\omega}t_{\omega},
		\label{Htr}
	\end{split}
\end{eqnarray}

\begin{eqnarray}
	\begin{split}
		\left \langle D_{(s)}^{+}K^{+}K^{-}\left | H^{P} \right | \ B_{c}^{+} \right \rangle=
		&
		\frac{g_{\phi\rightarrow K^{+} K^{-}}}{s_{\phi}}p_{\phi}
		+\frac{g_{\rho\rightarrow K^{+} K^{-}}}{s_{\rho}s_{\phi}}\widetilde{\Pi}_{\rho\phi}p_{\phi}
		+\frac{g_{\omega\rightarrow K^{+} K^{-}}}{s_{\omega}s_{\phi}}\widetilde{\Pi}_{\omega\phi}p_{\phi}
		+\frac{g_{\rho\rightarrow K^{+} K^{-}}}{s_{\rho}}p_{\rho}
		+\frac{g_{\phi\rightarrow K^{+} K^{-}}}{s_{\phi}s_{\rho}}\widetilde{\Pi}_{\phi\rho}p_{\rho}
		\\
		&
		+\frac{g_{\omega\rightarrow K^{+} K^{-}}}{s_{\omega s_{\rho}}}\widetilde{\Pi}_{\omega\rho}p_{\rho}
		+\frac{g_{\omega\rightarrow K^{+} K^{-}}}{s_{\omega}}p_{\omega}
		+\frac{g_{\phi\rightarrow K^{+} K^{-}}}{s_{\phi}s_{\omega}}\widetilde{\Pi}_{\phi\omega}p_{\omega}
		+\frac{g_{\rho\rightarrow K^{+} K^{-}}}{s_{\rho}s_{\omega}}\widetilde{\Pi}_{\rho\omega}p_{\omega}.
		\label{Hpe}
	\end{split}
\end{eqnarray}
Here, $g_{V}$ denotes the coupling constant obtained from the decay process $V \rightarrow K^{+} K^{-}$. Additionally, the following relationship holds: $\sqrt{2}g_{{\rho}K^{+} K^{-}} = \sqrt{2}g_{\omega K^{+} K^{-}} = -g_{\phi K^{+} K^{-}} = 4.54$ \cite{Bruch:2004py,Cheng:2020ipp}.

\subsection{Formulation of calculations}

In the decay process of $B_{c}^{+} \rightarrow D_{(s)}^{+}K^{+}K^-$, nine distinct scenarios are illustrated in Fig. 5. Among these, cases (b), (d), and (i) closely resemble the decay process of $B_{c}^{+} \rightarrow D_{(s)}^{+}\pi^{+}\pi^-$, differing only by the substitution of $\pi$ with K. The amplitude forms for these cases are similar to those presented in Eqs. (13)-(18). Focusing on the $B_{c}^{+} \rightarrow D^{+}K^{+}K^-$ decay process as an example, we outline all corresponding amplitude forms depicted in Fig. 5. These forms sequentially correspond to cases (a), (e), (h), (g), (c), and (f) in Fig. 5:  
\begin{equation}
	\begin{aligned}
		A\left(\ B_{c}^{+} \rightarrow\right.\left. D^{+}\phi\rightarrow D^{+}K^{+}K^{-}\right)=&\frac{<D^{+}\phi|H_eff|B_{c}^{+}><K^{+}K^{-}|H_{\phi \rightarrow K^{+}K^{-} }|\phi>}{({s-m_{\phi}^{2}+im_{\phi} \varGamma_{\phi}})} \\&
		=\sum_{\lambda =0,\pm 1} \frac{G_{F}P_{(B_{c}^{+}}\cdot \epsilon^*\left( \lambda \right)\ g^{\phi \rightarrow K^{+}K^{-} }\epsilon \left( \lambda \right) \cdot \left( p_{K ^+}-p_{K ^-} \right)}{\sqrt{2}({s-m_{\phi}^{2}+im_{\phi} \varGamma_{\phi}})}
		\\	&
		\times \bigg\{-V_{u d} V_{u b}^{*}\left[(C_4-\frac{1}{2}C_{10})\mathcal {M}_{e}^{LL}+(C_6-\frac{1}{2}C_8)\mathcal {M}_{e}^{SP}+(C_3+\frac{1}{3}C_4\right.\\    &\left.-\frac{1}{2}C_9-\frac{1}{6}C_{10}){F}_{e}^{LL}+(C_5+\frac{1}{3}C_6-\frac{1}{2}C_7-\frac{1}{6}C_8){F}_{e}^{LR}\right]\bigg\},
	\end{aligned}
\end{equation}

\begin{equation}
	\begin{aligned}
		\sqrt{2}A\left(\ B_{c}^{+} \rightarrow\right.\left. D^{+}(\rho^{0}-\phi)\rightarrow D^{+}K^{+}K^{-}\right) =&  \frac{<D^{+}\rho ^0|H_eff|B_{c}^{+}><K^{+}K^{-}|H_{\phi\rightarrow K^{+}K^{-} }|\phi>\widetilde{\Pi}_{\phi\rho}}{({s-m_{\rho ^0}^{2}+im_{\rho ^0} \varGamma _{\rho ^0}})({s-m_{\phi}^{2}+im_{\phi} \varGamma_{\phi}})}\\ & 
		=\sum_{\lambda =0,\pm 1} \frac{G_{F}P_{(B_{c}^{+}}\cdot \epsilon^*\left( \lambda \right)\ g^{\phi\rightarrow K^{+}K^{-} }\epsilon \left( \lambda \right) \cdot \left( p_{K ^+}-p_{K ^-} \right)\widetilde{\Pi}_{\phi\rho}}{\sqrt{2}{({s-m_{\rho ^0}^{2}+im_{\rho ^0} \varGamma _{\rho ^0}}})({s-m_{\phi}^{2}+im_{\phi} \varGamma_{\phi}})}
		\\& 
		\left.\times \Bigg\{V_{u d} V_{u b}^{*}\left[F_{e}^{L L}(C_1+\frac{1}{3}C_2)+M_{e}^{L L}(C_{2})+F_{a}^{L L }(C_{2}+\frac{1}{3}C_1)+M_{a}^{L L}(C_{1})\right]  \right.\\
		&+ V_{t d} V_{t b}^{*}\bigg\{F_{a}^{L L}(C_{2}+\frac{1}{3}C_1)+M_{a}^{L L}(C_{1})-\left[\mathcal {M}_{e}^{LL}(\frac{3}{2}C_{10}-C_3+\frac{1}{2}C_9)\right.\\
		&\left.-M_{a}^{L L}(C_{3}+C_{9})+M_{e}^{L R}(-C_{5}+\frac{1}{2}C_{7})+(-C_4-\frac{1}{3}C_3-C_{10}-\frac{1}{3}C_9){F}_a^{LL}\right.\\
		&\left.+(C_{10}+\frac{5}{3}C_9-\frac{1}{3}C_3-C_4-\frac{3}{2}C_7-\frac{1}{2}C_8){F}_e^{LL}\right.\\
		&\left.+(-C_6-\frac{1}{3}C_5+\frac{1}{2}C_{8}+\frac{1}{6}C_7){F}_e^{SP}-(C_5+C_7)\mathcal {M}_a^{LR}\right.\\
		&\left.+(-C_6-\frac{1}{3}C_5-C_{8}-\frac{1}{3}C_7) {F}_a^{SP}\right]\bigg\}\Bigg\},
	\end{aligned}
  \end{equation}

\begin{equation}
	\begin{aligned}
		\sqrt{2}A\left(\ B_{c}^{+} \rightarrow\right.\left. D^{+}(\omega-\phi)\rightarrow D^{+}K^{+}K^{-}\right)=&\frac{<D^{+}\omega|H_eff|B_{c}^{+}><K^{+}K^{-}|H_{\phi\rightarrow K^{+}K^{-} }|\phi>\widetilde{\Pi}_{\phi\omega}}{({s-m_{\omega}^{2}+im_{\omega} \varGamma_{\omega}})({s-m_{\phi}^{2}+im_{\phi} \varGamma_{\phi}})} \\&
		=\sum_{\lambda =0,\pm 1} \frac{G_{F}P_{(B_{c}^{+}}\cdot \epsilon^*\left( \lambda \right)\ g^{\phi\rightarrow K^{+}K^{-} }\epsilon \left( \lambda \right) \cdot \left( p_{K ^+}-p_{K ^-} \right)\widetilde{\Pi}_{\phi\omega}}{\sqrt{2}({s-m_{\omega}^{2}+im_{\omega} \varGamma_{\omega}})({s-m_{\phi}^{2}+im_{\phi} \varGamma_{\phi}})}\\	&
		\left.\times \bigg\{V_{u d} V_{u b}^{*}\left[ F_{e}^{L L}\left(c_{1}+\frac{1}{3}C_2\right)+C_2\mathcal {M}_{e}^{LL}-\left((C_2+\frac{1}{3}C_1)F_{a}^{L L}+C_1\mathcal {M}_{a}^{LL}\right)\right]\right.\\
		&\left.-V_{t d} V_{t b}^{*}\left[(C_2+\frac{1}{3}C_1)F_{a}^{L L}+C_1\mathcal {M}_{a}^{LL}+\left((2C_4+C_3+\frac{1}{2}C_{10}-\frac{1}{2}C_9)\mathcal {M}_{e}^{LL}\right)\right.\right.\\
		&\left.+(C_3+C_9)\mathcal {M}_{e}^{LL}+(C_5-\frac{1}{2}C_7)\mathcal {M}_{e}^{LR}+(C_5+C_7)\mathcal {M}_{a}^{LR}\right.\\
		&\left.+(C_4+\frac{1}{3}C_3+C_{10}+\frac{1}{3}C_9) {F}_{a}^{LL}+(\frac{7}{3}C_3+\frac{5}{3}C_4+\frac{1}{3}(C_9-C_{10})){F}_{e}^{LL}\right.\\
		&\left.+(2C_5+\frac{2}{3}C_6+\frac{1}{2}C_7+\frac{1}{6}C_8){F}_{e}^{LR}+(C_6+\frac{1}{3}C_5-\frac{1}{2}C_{8}-\frac{1}{6}C_7){F}_{e}^{SP}\right.\\
		&\left.+(C_6+\frac{1}{3}C_5+C_{8}+\frac{1}{3}C_7){F}_{a}^{SP}\right]\bigg\},
	\end{aligned}
\end{equation}

\begin{equation}
	\begin{aligned}
		\sqrt{2}A\left(\ B_{c}^{+} \rightarrow\right.\left. D^{+}\omega\rightarrow D^{+}K^{+}K^{-}\right)=&\frac{<D^{+}\omega|H_eff|B_{c}^{+}><K^{+}K^{-}|H_{\omega\rightarrow K^{+}K^{-} }|\omega>}{({s-m_{\omega}^{2}+im_{\omega} \varGamma_{\omega}})} \\&
		=\sum_{\lambda =0,\pm 1} \frac{G_{F}P_{(B_{c}^{+}}\cdot \epsilon^*\left( \lambda \right)\ g^{\omega\rightarrow K^{+}K^{-} }\epsilon \left( \lambda \right) \cdot \left( p_{K ^+}-p_{K ^-} \right)}{\sqrt{2}({s-m_{\omega}^{2}+im_{\omega} \varGamma_{\omega}})}\\	&
		\left.\times \bigg\{V_{u d} V_{u b}^{*}\left[ F_{e}^{L L}\left(c_{1}+\frac{1}{3}C_2\right)+C_2\mathcal {M}_{e}^{LL}-\left((C_2+\frac{1}{3}C_1)F_{a}^{L L}+C_1\mathcal {M}_{a}^{LL}\right)\right]\right.\\
		&\left.-V_{t d} V_{t b}^{*}\left[(C_2+\frac{1}{3}C_1)F_{a}^{L L}+C_1\mathcal {M}_{a}^{LL}+\left((2C_4+C_3+\frac{1}{2}C_{10}-\frac{1}{2}C_9)\mathcal {M}_{e}^{LL}\right)\right.\right.\\
		&\left.+(C_3+C_9)\mathcal {M}_{e}^{LL}+(C_5-\frac{1}{2}C_7)\mathcal {M}_{e}^{LR}+(C_5+C_7)\mathcal {M}_{a}^{LR}\right.\\
		&\left.+(C_4+\frac{1}{3}C_3+C_{10}+\frac{1}{3}C_9) {F}_{a}^{LL}+(\frac{7}{3}C_3+\frac{5}{3}C_4+\frac{1}{3}(C_9-C_{10})){F}_{e}^{LL}\right.\\
		&\left.+(2C_5+\frac{2}{3}C_6+\frac{1}{2}C_7+\frac{1}{6}C_8){F}_{e}^{LR}+(C_6+\frac{1}{3}C_5-\frac{1}{2}C_{8}-\frac{1}{6}C_7){F}_{e}^{SP}\right.\\
		&\left.+(C_6+\frac{1}{3}C_5+C_{8}+\frac{1}{3}C_7){F}_{a}^{SP}\right]\bigg\},
	\end{aligned}
\end{equation}

\begin{equation}
	\begin{aligned}
		A\left(\ B_{c}^{+} \rightarrow\right.\left. D^{+}(\phi-\omega)\rightarrow D^{+}K^{+}K^{-}\right)=&\frac{<D^{+}\phi|H_eff|B_{c}^{+}><K^{+}K^{-}|H_{\omega\rightarrow K^{+}K^{-} }|\omega>\widetilde{\Pi}_{\omega\phi}}{({s-m_{\phi}^{2}+im_{\phi} \varGamma_{\phi}})({s-m_{\omega}^{2}+im_{\omega} \varGamma_{\omega}})} \\&
		=\sum_{\lambda =0,\pm 1} \frac{G_{F}P_{(B_{c}^{+}}\cdot \epsilon^*\left( \lambda \right)\ g^{\omega\rightarrow K^{+}K^{-} }\epsilon \left( \lambda \right) \cdot \left( p_{K ^+}-p_{K ^-} \right)\widetilde{\Pi}_{\omega\phi}}{\sqrt{2}({s-m_{\phi}^{2}+im_{\phi} \varGamma_{\phi}})({s-m_{\omega}^{2}+im_{\omega} \varGamma_{\omega}})}
		\\	&
		\times \bigg\{-V_{u d} V_{u b}^{*}\left[(C_4-\frac{1}{2}C_{10})\mathcal {M}_{e}^{LL}+(C_6-\frac{1}{2}C_8)\mathcal {M}_{e}^{SP}+(C_3+\frac{1}{3}C_4\right.\\    &\left.-\frac{1}{2}C_9-\frac{1}{6}C_{10}){F}_{e}^{LL}+(C_5+\frac{1}{3}C_6-\frac{1}{2}C_7-\frac{1}{6}C_8){F}_{e}^{LR}\right]\bigg\},
	\end{aligned}
\end{equation}
and
\begin{equation}
	\begin{aligned}
		\sqrt{2}A\left(\ B_{c}^{+} \rightarrow\right.\left. D^{+}(\rho^{0}-\omega)\rightarrow D^{+}K^{+}K^{-}\right) =&  \frac{<D^{+}\rho ^0|H_eff|B_{c}^{+}><K^{+}K^{-}|H_{\omega\rightarrow K^{+}K^{-}}|\omega>\widetilde{\Pi}_{\omega\rho}}{({s-m_{\rho ^0}^{2}+im_{\rho ^0} \varGamma _{\rho ^0}})({s-m_{\omega}^{2}+im_{\omega} \varGamma_{\omega}})}\\ & 
		=\sum_{\lambda =0,\pm 1} \frac{G_{F}P_{(B_{c}^{+}}\cdot \epsilon^*\left( \lambda \right)\ g^{\omega\rightarrow K^{+}K^{-} }\epsilon \left( \lambda \right) \cdot \left( p_{K ^+}-p_{K ^-} \right)\widetilde{\Pi}_{\omega\rho}}{\sqrt{2}{({s-m_{\rho ^0}^{2}+im_{\rho ^0} \varGamma _{\rho ^0}}})({s-m_{\omega}^{2}+im_{\omega} \varGamma_{\omega}})}
		\\& 
		\left.\times \Bigg\{V_{u d} V_{u b}^{*}\left[F_{e}^{L L}(C_1+\frac{1}{3}C_2)+M_{e}^{L L}(C_{2})+F_{a}^{L L }(C_{2}+\frac{1}{3}C_1)+M_{a}^{L L}(C_{1})\right]  \right.\\
		&+ V_{t d} V_{t b}^{*}\bigg\{F_{a}^{L L}(C_{2}+\frac{1}{3}C_1)+M_{a}^{L L}(C_{1})-\left[\mathcal {M}_{e}^{LL}(\frac{3}{2}C_{10}-C_3+\frac{1}{2}C_9)\right.\\
		&\left.-M_{a}^{L L}(C_{3}+C_{9})+M_{e}^{L R}(-C_{5}+\frac{1}{2}C_{7})+(-C_4-\frac{1}{3}C_3-C_{10}-\frac{1}{3}C_9){F}_a^{LL}\right.\\
		&\left.+(C_{10}+\frac{5}{3}C_9-\frac{1}{3}C_3-C_4-\frac{3}{2}C_7-\frac{1}{2}C_8){F}_e^{LL}\right.\\
		&\left.+(-C_6-\frac{1}{3}C_5+\frac{1}{2}C_{8}+\frac{1}{6}C_7){F}_e^{SP}-(C_5+C_7)\mathcal {M}_a^{LR}\right.\\
		&\left.+(-C_6-\frac{1}{3}C_5-C_{8}-\frac{1}{3}C_7) {F}_a^{SP}\right]\bigg\}\Bigg\}.
	\end{aligned}
\end{equation}

\subsection{CP violation results of the $\ B_{c}^{+} \rightarrow   D_{(s)}^{+}K^{+}K^-$  decay process}

\begin{figure}[h]
	\centering
	
	\includegraphics[height=6cm,width=9cm]{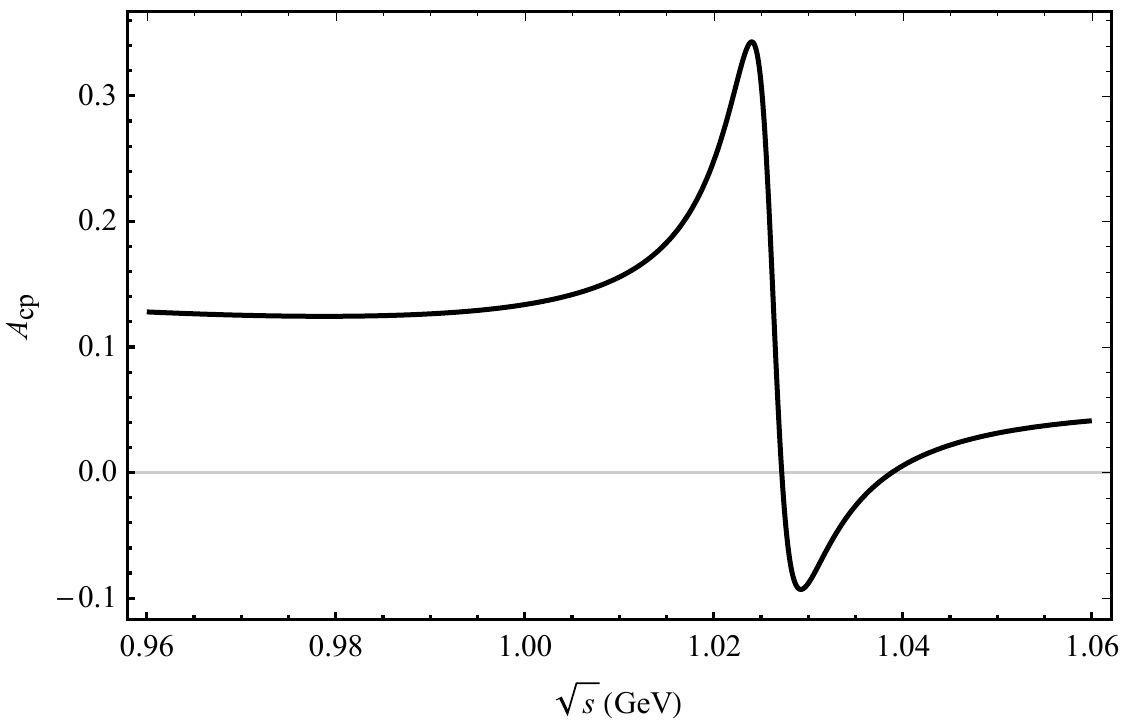}
	
	\caption{Plot of  $A_{CP}$ as a function of $\sqrt{s}$ corresponding to central parameter values of CKM matrix elements
		for the decay channel of $ B_{c}^{+} \rightarrow D^{+}K^{+}K^{-}$.}
	\label{fig:3}
\end{figure}

\begin{figure}[h]
	\centering
	
	\includegraphics[height=6cm,width=9cm]{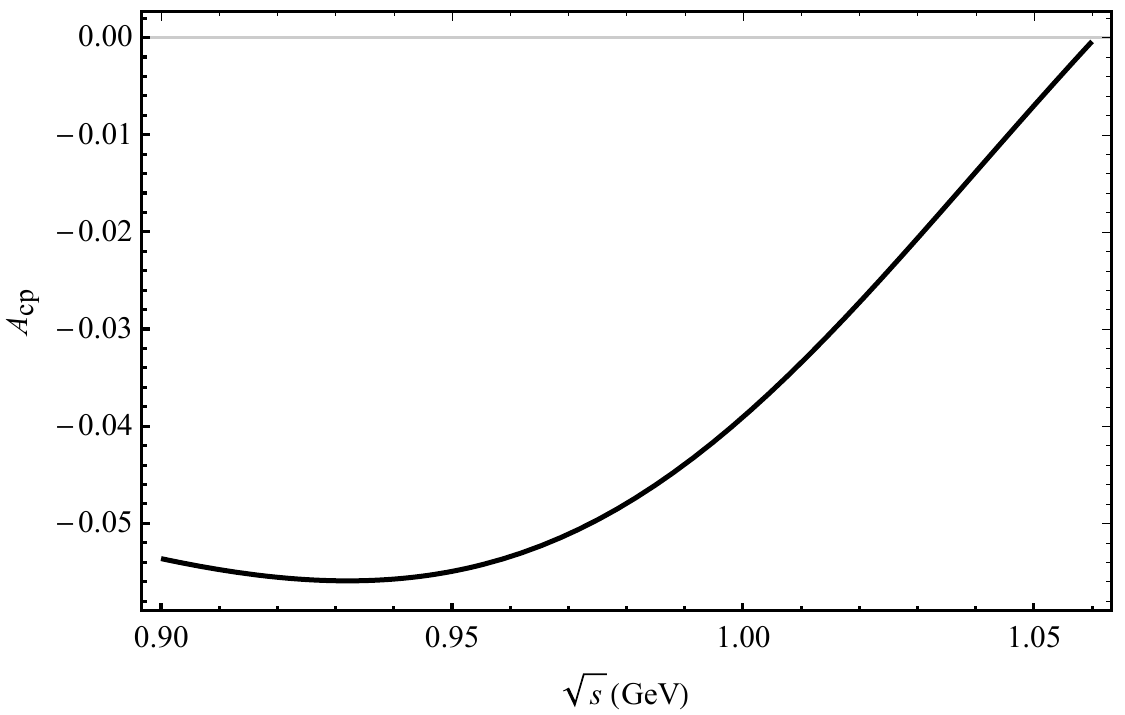}
	
	\caption{Plot of  $A_{CP}$ as a function of $\sqrt{s}$ corresponding to central parameter values of CKM matrix elements
		for the decay channel of $ B_{c}^{+} \rightarrow D_s^{+}K^{+}K^{-}$.}
	\label{fig:3}
\end{figure}

We present plots illustrating CP violation in the decay processes of $ B_{c}^{+} \rightarrow D^{+}K^{+}K^{-}$ and $ B_{c}^{+} \rightarrow D_s^{+}K^{+}K^{-}$. Taking into account the $K^{+}K^{-}$ threshold, Fig. 6 and 7 depict the variation of $A_{CP}$ as a function of $\sqrt{s}$, representing the invariant mass of the $K^{+}K^{-}$ system. In the decay process of $ B_{c}^{+} \rightarrow D^{+}K^{+}K^{-}$ shown in Fig. 6, a pronounced CP-violating effect is observed near 1.02 GeV. The peak value reaches $34.30\%$ when the invariant mass of $K^+K^-$ is close to the mass of the $\phi$ meson (1.02 GeV).
In the decay process of $B_{c}^{+} \rightarrow D^{+}\phi$ associated with the $B_{c}^{+} \rightarrow D^{+}K^{+}K^{-}$ decay, it is observed that only the penguin diagram contributes, while there is no contribution from the tree diagram, which does not induce CP violation. The CP violation depends on the intermediate decay processes $\rho \rightarrow K^{+}K^{-}$ and $\omega \rightarrow K^{+}K^{-}$, as well as the interference between these processes and $\phi \rightarrow K^{+}K^{-}$.

In the decay process of $ B_{c}^{+} \rightarrow D_s^{+}K^{+}K^{-}$ as shown in Fig. 7, an intriguing phenomenon has been observed. The CP violation exhibits a sharp increase when the invariant mass of the $K^+K^-$ pair is around 0.93 GeV. Notably, this peak does not align with the mass of the $\phi$ meson; instead, it spans a broader range from 0.9 to 1.05 GeV. We propose that this behavior results from the resonant mixing of the $\rho^0$, $\omega$, and $\phi$ particles, with the decay channel $\phi \rightarrow K^{+}K^{-}$ making the dominant contribution through their combined effect.

\section{The localised CP violation  of $A^{\Omega}_{CP}$}

In this paper, we perform the integral calculation of A$_{CP}$ to facilitate future experimental comparisons. For the decay process$\ B_{c}^{+} \rightarrow D_{(s)}^{+}\rho^0$, the amplitude is given by $M_{\ B_{c}^{+}\rightarrow D_{(s)}^{+}\rho^0}^{\lambda}=\alpha p_{B_{c}^{+}} \cdot \epsilon^{*}(\lambda)$, where $p_{B_{c}^{+}}$ represents the momenta of the$\ B_{c}^{+}$ meson, $\epsilon$ denotes the polarization vector of $\rho^0$ and $\lambda$ corresponds to its polarization. The parameter $\alpha$ remains independent of $\lambda$
which is from the contribution of PQCD. Similarly, in the decay process $\rho^0 \rightarrow \pi^{+}\pi^{-}$, we can express $M_{\rho^0 \rightarrow \pi^{+}\pi^{-}}^{\lambda}=g_{\rho}\epsilon(\lambda)\left(p_1-p_2\right)$, where $p_1$ and $p_2$ denote the momenta of the produced $\pi^{+}$ and $\pi^{-}$ particles from $\rho^0$ meson, respectively. Here, the parameter $g_\rho$ represents an effective coupling constant for $\rho^0 \rightarrow \pi^{+}\pi^{-}$. Regarding the dynamics of meson decay, it is observed that the polarization vector of a vector meson satisfies $\sum_{\lambda=0,\pm 1}\epsilon^\lambda_\mu(p)(\epsilon^\lambda_\nu(p))^*=-(g_{\mu\nu}-p_\mu p_\nu/m_V^2)$. As a result, we obtain the total amplitude for the decay process $\ B_{c}^{+} \rightarrow  D_{(s)}^{+}\rho^0\rightarrow D_{(s)}^{+}\pi^{+}\pi^-$
\cite{Rui:2011qc}:
\begin{equation}
	\begin{aligned}
		A &=\alpha p_{\ B_{c}^{+}}^{\mu} \frac{\sum_{\lambda} \epsilon_{\mu}^{*}(\lambda) \epsilon_{\nu}(\lambda)}{s_{\rho}} g_{\rho \pi\pi}\left(p_{1}-p_{2}\right)^{\nu} \\
		&=\frac{g_{\rho \pi\pi} \alpha}{s_{\rho}} \cdot p_{\ B_{c}^{+}}^{\mu}\left[g_{\mu \nu}-\frac{\left(p_{1}+p_{2}\right)_{\mu}\left(p_{1}+p_{2}\right)_{\nu}}{s}\right]\left(p_{1}-p_{2}\right)^{\nu} \\
		&=\frac{g_{\rho \pi\pi}}{s_{\rho}} \cdot \frac{M_{\ B_{c}^{+}\rightarrow \rho^0 D_{(s)}^{+}}^{\lambda}}{p_{\ B_{c}^{+}} \cdot \epsilon^{*}} \cdot\left(\Sigma-s^{\prime}\right) \\
		&=\left(\Sigma-s^{\prime}\right) \cdot \mathcal{A}.
	\end{aligned}
\end{equation}

The high ($\sqrt{s^\prime}$) and low $\sqrt{s}$ ranges are defined for calculating the invariant mass of $\pi^{+} \pi^{-}$. By setting a fixed value for $s$, we can determine an appropriate value for $s^\prime$ that fulfills the equation $\Sigma=\frac{1}{2}\left(s_{\max }^\prime+s_{\min }^\prime\right)$, where  ${ s}_{ \max  }^{ \prime  }( { s}_{ \min  }^{\prime}) $ denotes the maximum (minimum) value, respectively.

Utilizing the principles of three-body kinematics, we can deduce the local CP asymmetry for the decay $\ B_{c}^{+} \rightarrow D_{(s)}^{+}\pi^{+}\pi^- $ within a specific range of invariant mass:
\begin{equation}
	A_{C P}^{\Omega}=\frac{\int_{s_{1}}^{s_{2}} \mathrm{~d} s \int_{s_{1}^{\prime}}^{s_{2}^{\prime}} \mathrm{d} s^{\prime}\left(\Sigma-s^{\prime}\right)^{2}\left(|\mathcal{A}|^{2}-|\overline{\mathcal{A}}|^{2}\right)}{\int_{s_{1}}^{s_{2}} \mathrm{~d} s \int_{s_{1}^{\prime}}^{s_{2}^{\prime}} \mathrm{d} s^{\prime}\left(\Sigma-s^{\prime}\right)^{2}\left(|\mathcal{A}|^{2}+|\overline{\mathcal{A}}|^{2}\right)}.
\end{equation}

Our calculation takes into account the dependence of $\Sigma=\frac{1}{2}\left(s_{\max }^{\prime}+s_{\min }^{\prime}\right)$ on $s^{\prime}$. Assuming that $s_{\max }^{\prime}>s^{\prime}>s_{\min }^{\prime}$ represents an integral interval of high invariant mass for the $\pi^{+} \pi^{-}$ meson pair, and $\int_{s_{1}^{\prime}}^{s_{2}^{\prime}} \mathrm{d}s^\prime(\Sigma-s')^{2}$ represents a factor dependent on $s'$. The correlation between $\Sigma$ and $s'$ can be easily determined through kinematic analysis, as $s'$ only varies on a small scale. Therefore, we can consider $\Sigma$ as a constant. This allows us to cancel out the term $\int_{s_1^\prime}^{ s_2^\prime }\mathrm{d}s^\prime (\Sigma-s')^{2}$ in both the numerator and denominator, resulting in $A_{C P}^{\Omega}$ no longer depending on the high invariant mass of positive and negative particles.
The form of $\ B_{c}^{+} \rightarrow D_{(s)}^{+}\phi$ ($\rho^0$, $\omega$) $   \rightarrow  D_{(s)}^{+}K^{+}K^{-}$ is similar to that of $\ B_{c}^{+} \rightarrow D_{(s)}^{+}\rho^{0}$ ($\omega$, $\phi$) $   \rightarrow  D_{(s)}^{+}\pi^{+}\pi^{-}$, only $ \pi\ $needs to be replaced with K.

\section{Numerical results of the localized integrated CP asymmetry}

\begin{table}[h]
	{\renewcommand
		\scalebox{15}
		\centering %
		\renewcommand{\arraystretch}{4} %
		\setlength{\tabcolsep}{2mm}{
			\begin{center}
				\caption{The peak local ($0.65$ $GeV\leq \sqrt{s}\leq 1.06$ $GeV$ for the $\pi^{+}\pi^{-}$
					final states and $0.98$ $GeV\leq \sqrt{s}\leq1.06$ $GeV$ for the $K^{+}K^{-}$
					final states					
					) integral of  $\mathrm{A}^{\Omega} _{\mathrm{CP}}$ from different resonance rangs for $\ B_{c}^{+} \rightarrow   D_{(s)}^{+}\pi^{+}\pi^-$ and $\ B_{c}^{+} \rightarrow   D_{(s)}^{+}K^{+}K^-$ decay processes.}
				\begin{tabular}{ ccccc  }
					\hline
					Decay channel                                     &$ \ B_{c}^{+} \rightarrow D^{+}\pi^{+}\pi^{-}$               &  $ B_{c}^{+} \rightarrow D_s^{+}\pi^{+}\pi^{-}$             & $ B_{c}^{+} \rightarrow D^{+}K^{+}K^{-}$             & $ B_{c}^{+} \rightarrow D_s^{+}K^{+}K^{-}$ \\ \hline
					$\phi-\rho-\omega$ mixing       & $\mathrm{0.0706^{+0.002-0.063}_{-0.002-0.091} }$             &$\mathrm{-0.0134^{+0.0005-0.0016}_{-0.0005+0.0006} }  $            &$ \mathrm{0.1636^{+0.006+0.046}_{-0.006-0.067} }$            &$ \mathrm{-0.0278^{+0.001-0.0004}_{-0.001+0.0012} }$ \\
					$\rho-\omega$ mixing       & $\mathrm{0.0706^{+0.002-0.064}_{-0.002-0.091} } $          &$\mathrm{-0.0133^{+0.0005-0.0013}_{-0.0005+0.0012} } $          &$ \mathrm{0.0906^{+0.006+0.074}_{-0.006-0.072} }$            &$ \mathrm{-0.0267^{+0.001+0.014}_{-0.001-0.004} }$    \\
					$\phi-\rho$ mixing       & $\mathrm{-0.0176^{+0.0006+0}_{-0.0006-0} }$          & $\mathrm{-0.0156^{+0.0006-0.0005}_{-0.0006+0.00001} }$         & $\mathrm{0.0290^{+0.0005-0.045}_{-0.0005-0.041} }$            &$ \mathrm{-0.0258^{+0.0009+0.0012}_{-0.0009+0.0008} }$    \\
					$\phi-\omega$ mixing     & $ \mathrm{-}$           & $\mathrm{-}$       &$\mathrm{-0.0381^{+0.0006+0.012}_{-0.0006-0.0096} }$            &$ \mathrm{-0.0264^{+0.00003-0.002}_{-0.00003-0.001} }$    \\
					no mixing     & $ \mathrm{-0.0176^{+0.0006+0}_{-0.0006-0} }$           & $\mathrm{-0.0155^{+0.0006+0}_{-0.0006-0} }$       &$\mathrm{0.0784^{+0.001+0.091}_{-0.001+0.050} }$            &$ \mathrm{-0.0252^{+0.0009+0.0015}_{-0.0009+0.0005} }$    \\
					\hline
				\end{tabular}
	\end{center}}}
\end{table}

According to Table I, the integration ranges of 0.65 GeV to 1.06 GeV and 0.98 GeV to 1.06 GeV correspond to the resonance regions for the $\pi^{+}\pi^{-}$ and $K^{+}K^{-}$ final states, respectively. The threshold for the $K^{+}K^{-}$ final state is also considered. The resonance interactions between different particles can result in more pronounced CP violation phenomena across various energy intervals. We present the local integral values as detailed in Table I. We have comparatively analyzed the distinctions between the three-particle mixed resonance of $\rho^0 -\omega-\phi$ and the two-particle mixed resonances of $\rho^0 -\omega$, $\phi-\rho^0$ and $\phi-\omega$, while also providing numerical results that exclude these mixed resonances.

During the decay process of $B_{c}^{+} \rightarrow D^{+}\pi^{+}\pi^{-}$, both in the case of a three-particle mixture and a $\rho^0 - \omega$ mixture, the variation of CP violation is significantly larger in the resonant region compared to the non-resonant scenario, with changes in sign. Despite a peak reaching $97.54\%$ of CP violation, the local integral value exhibits only minor fluctuations, and the central value of $\mathrm{A}^{\Omega}_{\mathrm{CP}}$ is approximately 0.0706. The influence of $\phi-\rho^0$ mixing is relatively insignificant compared to the no-mixing results. Additionally, the effect of $\phi-\omega$ mixing, arising from higher-order contributions with very small values, can be neglected. Therefore, we do not present these results.
For the decay processes of $B_{c}^{+} \rightarrow D_s^{+}\pi^{+}\pi^{-}$, the three-particle mixing mechanism exhibits negligible effects on CP violation. In the decay processes of both $B_{c}^{+} \rightarrow D^{+}\pi^{+}\pi^{-}$ and $B_{c}^{+} \rightarrow D_s^{+}\pi^{+}\pi^{-}$, the values of the $\rho^0 - \omega - \phi$ mixing and the $\rho^0 - \omega$ mixing are essentially identical. The influence of $\phi - \rho^0$ mixing is relatively minor and can be disregarded.

The effect of the mixing mechanism in the decay process of $ B_{c}^{+} \rightarrow D^{+}K^{+}K^{-}$ is significant, with the central value of $\mathrm{A}^{\Omega}_{\mathrm{CP}}$ reaching 0.1636 for CP violation, compared to 0.0784 without mixing. This is because in the decay process of $ B_{c}^{+} \rightarrow D^{+}\phi$, only the penguin diagram contributes, while there is no contribution from the tree diagram. Consequently, considering the three-particle mixing, the change in CP violation becomes more pronounced. However, in the process $ B_{c}^{+} \rightarrow D_s^{+}K^{+}K^{-}$, the CP violation exhibits a notable increase due to the three-particle mixing.

 Theoretical errors give rise to uncertainties in the results. In general, the major theoretical uncertainties arise from power corrections beyond the heavy quark limit, necessitating the inclusion of $1/m_b$ power corrections. However, the $1/m_b$ corrections are power-suppressed and typically non-perturbative, meaning they cannot be accurately calculated using perturbation theory. Consequently, this scheme introduces additional sources of uncertainty. The first source of error comes from variations in CKM parameters, while the second source of error comes from hadronic parameters, such as mixing parameters, form factors, decay constants, $B_c$ meson wave functions, and $D$ meson wave functions. By employing central values for these parameters, we initially compute numerical results for CP violation and subsequently incorporate errors based on standard deviations as shown in Table I.

\section{The branching ratio of $\ B_{c}^{+} \rightarrow D_{(s)}^{+}V \rightarrow  D_{(s)}^{+}K^{+}K^{-}$ }
\subsection{The branching ratios of different decay channels  }

Due to isospin breaking, the effects of three-particle mixing on the branching ratios of $B_{c}^{+} \rightarrow D_{(s)}^{+}\rho^{0} \rightarrow D_{(s)}^{+}\pi^{+}\pi^{-}$ are symmetrical. By considering $\sqrt{2}g_{{\rho}K^{+} K^{-}} = \sqrt{2}g_{\omega K^{+} K^{-}} = -g_{\phi K^{+} K^{-}} = 4.54$, we calculate the branching ratios of $B_{c}^{+} \rightarrow D_{(s)}^{+}V \rightarrow D_{(s)}^{+}K^{+}K^{-}$ both with and without three-particle mixing.
The differential branching ratios for the quasi-two-body $\ B_{c}^{+} \rightarrow D_{(s)}^{+}V \rightarrow  D_{(s)}^{+}K^{+}K^{-}$ 
decays is written as \cite{Chai:2021kie}:
\begin{equation}
\frac{d{\mathcal B}}{d\zeta} = \frac{\tau_{B_{c}} \, q^3_{D_{(s)}^{+}} \, q^3}{48 \, \pi^3 \, m^5_{B_{c}}} \, \overline{\vert {\mathcal A} \vert^2} \,,
\label{eq:diff-bra}
\end{equation}

with the variable ${\zeta}=\frac{s}{m_{B_{c}}^2}$
and the $B_c$ meson mean lifetime $\tau_{B_{c}}$.
Among them, $q$ and $q_{D_{(s)}^{+}}$ are respectively defined as 
\begin{equation}
q =\frac{1}{2} \sqrt{ s-(m_K+m_K)^2},
\end{equation}
and
\begin{equation}
q_{D_{(s)}^{+}}=\frac{1}{2}\sqrt{\big[\left(m^2_{B}-m_{D_{(s)}^{+}}^2\right)^2 -2\left(m^2_{B}+m_{D_{(s)}^{+}}^2\right)s+s^2\big]/s} \,. 
\end{equation}

In calculating the decays of $\ B_{c}^{+} \rightarrow D_{(s)}^{+}\rho^0 \rightarrow  D_{(s)}^{+}K^{+}K^{-}$ and $\ B_{c}^{+} \rightarrow D_{(s)}^{+}\omega \rightarrow  D_{(s)}^{+}K^{+}K^{-}$, the $\rho^0$ and $\omega$ pole masses are below the invariant mass threshold, i.e., $m_\rho(\omega) < m_K + m_K$\cite{Wang:2020nel}. In this case, the pole mass of $m_\rho(\omega)$ should be replaced by an effective mass $m_0^{\text{eff}}$, to avoid kinematic singularities in the phase space factor \cite{LHCb:2015eqv}:
\begin{equation}
{\ m_0^{\text{eff}}(m_0)=m^{min}+(m^{max}-m^{min})\times \left[1+\tanh \left( \frac{m_0-(m^{max}+m^{min})/2}{m^{max}-m^{min}} \right) \right]},
\end{equation}
here ${\rm m^{max}} = m_{B_{c}} - m_{D_{(s)}}$ and ${\rm m^{min}} = m_{K} + m_K $ 
are the upper and lower thresholds of $\sqrt{s}$, respectively.

\subsection{The results of the branching ratio in the decay process of $\ B_{c}^{+} \rightarrow D_{(s)}^{+}V \rightarrow  D_{(s)}^{+}K^{+}K^{-}$}

\begin{figure}[h]
	\centering
	\begin{minipage}[h]{0.45\textwidth}
		\centering
		\includegraphics[height=4cm,width=6.5cm]{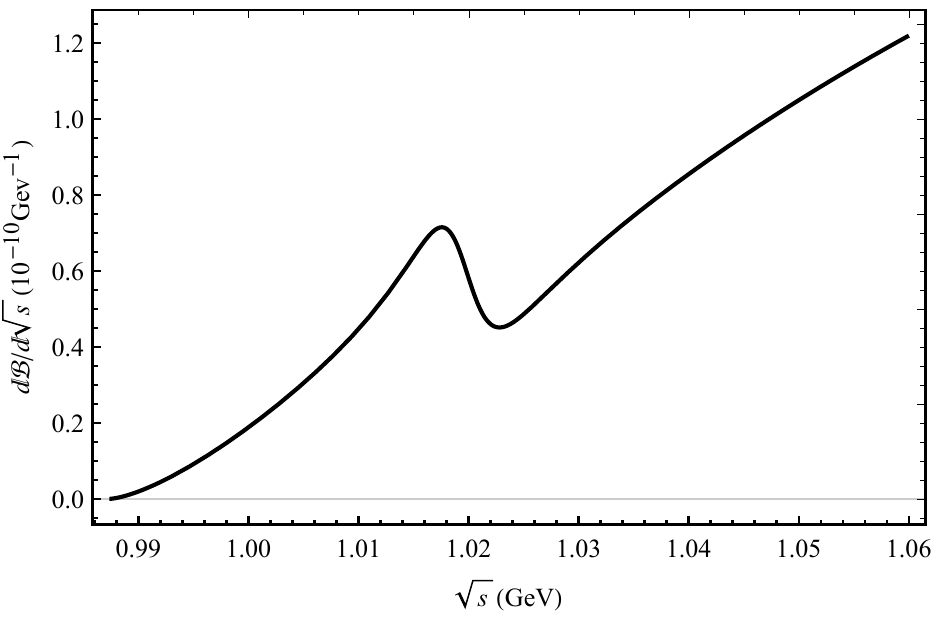}
		\caption{The differential branching ratios for the  $\ B_{c}^{+} \rightarrow D^{+}V \rightarrow  D^{+}K^{+}K^{-}$ decay process from the mixing of $\phi$, $\rho$, and $\omega$ mesons.}
		\label{fig2}
	\end{minipage}
	\quad
	\begin{minipage}[h]{0.45\textwidth}
		\centering
		\includegraphics[height=4cm,width=6.5cm]{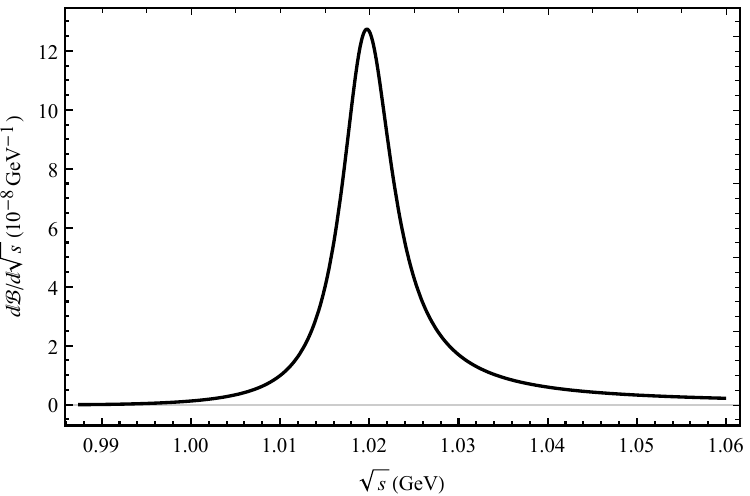}
		\caption{The differential branching ratios for the  $\ B_{c}^{+} \rightarrow D_{s}^{+}V \rightarrow  D_{s}^{+}K^{+}K^{-}$ decay process from the mixing of $\phi$, $\rho$, and $\omega$ mesons.}
		\label{fig3}
	\end{minipage}
\end{figure}

We present the result graphs of the branching ratio for the decay process $B_{c}^{+} \rightarrow D_{(s)}^{+}V \rightarrow D_{(s)}^{+}K^{+}K^{-}$, arising from the mixing of $\phi$, $\rho$, and $\omega$ mesons, in Fig. 8 and Fig. 9, respectively. 
One can observe the branching ratio of $B_{c}^{+} \rightarrow D_{(s)}^{+}V \rightarrow D^{+}K^{+}K^{-}$ exhibiting a rise and fall near the mass of the $\phi$ meson in Fig. 8. A prominent peak in the branching ratio is observed near the mass of the $\phi$ meson for the decay channel $B_{c}^{+} \rightarrow D_{s}^{+}V \rightarrow D_{s}^{+}K^{+}K^{-}$ in Fig. 9. Furthermore, we have computed the local integrals of the branching ratios, which can be directly compared with experimental data in Table II.

\begin{table}[h]
	\centering
	\renewcommand{\arraystretch}{2}
	\captionsetup{justification=centering}
	\caption{The PQCD predictions of the branching ratios for quasi-two-body decays $\ B_{c}^{+} \rightarrow D^{+}V \rightarrow  D^{+}K^{+}K^{-}$ and $\ B_{c}^{+} \rightarrow D_s^{+}V \rightarrow  D_s^{+}K^{+}K^{-}$.}
	\label{tab:my_table}
		\begin{supertabular}{>{\centering\arraybackslash}p{0.33\textwidth}>{\centering\arraybackslash}p{0.33\textwidth}>{\centering\arraybackslash}p{0.33\textwidth}}
			\toprule
			\textbf{Decay channel}                                     & \textbf{Branching ratio}        & \textbf{Direct three-body decay/Mixed three-body decay} \\ 
			\hline	\hline
			\midrule
			$\ B_{c}^{+} \rightarrow D^{+}\rho^0 \rightarrow  D^{+}K^{+}K^{-}$       & $\mathrm{1.66^{+0.15}_{-0.31}\times10^{-11}}$        & 35.37\%   \\ 
			$\ B_{c}^{+} \rightarrow D^{+}\omega \rightarrow  D^{+}K^{+}K^{-}$       & $\mathrm{2.36^{+0.37 }_{-0.37 }\times10^{-11}} $         & 50.21\%  \\ 
			$\ B_{c}^{+} \rightarrow D^{+}\phi \rightarrow  D^{+}K^{+}K^{-}$       & $\mathrm{4.37^{+0.20}_{-0.21 }\times10^{-11}}  $        & 92.98\%  \\ 
			$\ B_{c}^{+} \rightarrow D^{+}\phi (\rho^0,\omega) \rightarrow  D^{+}K^{+}K^{-}$       &
			 $\mathrm{4.70^{+0.53}_{-0.51 }\times10^{-11}}  $        & 1 \\ 
			 	\hline
			$\ B_{c}^{+} \rightarrow D_s^{+}\rho^0 \rightarrow  D_s^{+}K^{+}K^{-}$     & $ \mathrm{1.10^{+0.15}_{-0.14 }\times10^{-11}}$          & 0.0083\%  \\ 
			$\ B_{c}^{+} \rightarrow D_s^{+}\omega \rightarrow  D_s^{+}K^{+}K^{-}$     & $ \mathrm{3.76^{+0.48}_{-0.47 }\times10^{-12}}$          & 0.0028\%  \\ 
			$\ B_{c}^{+} \rightarrow D_s^{+}\phi \rightarrow  D_s^{+}K^{+}K^{-}$     & $ \mathrm{1.44^{+0.95}_{-0.24 }\times10^{-7}}$          & 109.09\% \\ 
			$\ B_{c}^{+} \rightarrow D_s^{+}\phi (\rho^0,\omega) \rightarrow  D_s^{+}K^{+}K^{-}$     & $ \mathrm{1.32^{+0.12}_{-0.12 }\times10^{-7}}$          & 1  \\
		\hline	\hline
			\bottomrule
		\end{supertabular}
\end{table}

We find that in the decay of $\ B_{c}^{+} \rightarrow D^{+}V \rightarrow  D^{+}K^{+}K^{-}$, $D^{+}\phi$ constitutes the dominant contribution, with a ratio for branching ratio of $92\%$, while the branching ratios of the three direct decay modes are of the same order of magnitude from 
$\ B_{c}^{+} \rightarrow D^{+}\phi\rightarrow  D^{+}K^{+}K^{-}$,
$\ B_{c}^{+} \rightarrow D^{+}\rho^0 \rightarrow  D^{+}K^{+}K^{-}$ and
	$\ B_{c}^{+} \rightarrow D^{+}\omega \rightarrow  D^{+}K^{+}K^{-}$.
The absence of tree-level diagrams in the two-body decay of $\ B_{c}^{+} \rightarrow D^{+}\phi$ leads to a small branching ratio for $\ B_{c}^{+} \rightarrow D^{+}\phi \rightarrow  D^{+}K^{+}K^{-}$. Meanwhile, the sum of the branching ratios of the three direct decays is significantly larger than the branching ratio of the mixed decay, which can be attributed to destructive interference.

In the decay of $\ B_{c}^{+} \rightarrow D_s^{+}V \rightarrow  D_s^{+}K^{+}K^{-}$, the branching ratio of $\ B_{c}^{+} \rightarrow D_s^{+}\phi \rightarrow  D_s^{+}K^{+}K^{-}$ constitutes $109\%$ of the branching ratio of the mixed decay. The branching ratios of the other two direct decays are four or five orders of magnitude lower than that of the mixed decay and are therefore negligible. The fact that the branching ratio of $\ B_{c}^{+} \rightarrow D_s^{+}\phi \rightarrow  D_s^{+}K^{+}K^{-}$ exceeds that of the mixed decay is also attributed to destructive interference.

\section{Summary and conclusion}

The CP violation in the decay process of $\ B_{c}^{+}$ meson is predicted through an invariant mass analysis of $\pi^+\pi^-$  and $K^{+} K^{-}$meson pairs within the resonance region, resulting from the mixing of $\phi$, $\omega$, and $\rho$ mesons. We observe a sharp change in CP violation within the resonance regions of these mesons. Local CP violation is quantified by integrating over phase space. For the decay process $ \ B_{c}^{+} \rightarrow D^{+}\pi^{+}\pi^{-}$ and  $ B_{c}^{+} \rightarrow D^{+}K^{+}K^{-}$, the CP violation change large from
the effects of  $\rho^0 -\omega-\phi$ mixing
at the ranges of resonance. 
Experimental detection of local CP violation can be achieved by reconstructing the resonant states of $\phi$, $\omega$, and $\rho$ mesons within the resonance regions.

Our calculation results for the branching ratios indicate that the combined branching ratios of the three direct decays exceed those of the mixed decays in the processes $B_{c}^{+} \rightarrow D^{+}\phi (\rho^0,\omega) \rightarrow D^{+}K^{+}K^{-}$ and $B_{c}^{+} \rightarrow D_s^{+}\phi (\rho^0,\omega) \rightarrow D_s^{+}K^{+}K^{-}$. The contributions from the three direct decays, along with additional interference terms, influence the overall branching ratio, especial for the decay process of $B_{c}^{+} \rightarrow D^{+}\phi (\rho^0,\omega) \rightarrow D^{+}K^{+}K^{-}$ .

 \section*{Acknowledgements}
 This work was supported by  Natural Science Foundation of Henan (Project no. 232300420115), National Natural Science Foundation of China (Project no. 12275024) and Cultivation Project of Tuoxin in Henan University of Technology.

\section*{\label{Appendix}  Appendix: Input parameters}

The $V_{t b}$, $V_{t s}$, $V_{u b}$, $ V_{u s}$, $ V_{t d}$, and $ V_{u d}$ terms in the above equation are derived from the CKM matrix element within the framework of the Standard Model.
The CKM matrix, whose elements are determined through experimental observations, can be expressed in terms of the Wolfenstein parameters $A$,
$\rho$, $\lambda$, and $\eta$:
$V_{t b} V_{t s}^{*}=\lambda$, $V_{u b} V_{u s}^{*}=A \lambda^{4}(\rho-i \eta)$, $V_{u b} V_{u d}^{*}=A \lambda^{3}(\rho-i \eta)(1-\frac{\lambda^{2}}{2})$, $V_{t b} V_{t d}^{*}=A \lambda^{3}(1-\rho+i \eta)$.
The most recent values for the parameters in the CKM matrix are $\lambda=0.22650\pm 0.00048$, $A=0.790_{-0.012}^{+0.017}$, $\bar{\rho}=0.141_{-0.017}^{+0.016}$, and $ \bar{\eta}=0.357\pm 0.011$. Here, we define $\bar{\rho}=\rho\left(1-\frac{\lambda^{2}}{2}\right)$ and $ \bar{\eta}=\eta\left(1-\frac{\lambda^{2}}{2}\right)$\cite{Wolfenstein:1983yz}
\cite{LHCb:2022fpg}.
The physical quantities involved in the calculation are presented in the Table III:
\begin{table}[h]
	\centering
	\caption{The input parameters  (in the unit of GeV)\cite{ParticleDataGroup:2024cfk, Wolfenstein:1964ks}}
	\sisetup{
		table-format=1.5e-1,
		table-number-alignment = center,
		exponent-product = \cdot,
	}
	\renewcommand{\arraystretch}{1.5}
	\begin{tabular*}{\textwidth}{@{\extracolsep{\fill}}llll}
		\hline
		\hline 
		\( m_{B_{c}} = \num{6.27447}\pm0.27 \) & \( f_{\phi} = \num{0.231} \) & \( m_{D_{s}^+} = \num{1.96835}\pm0.07 \) & \( f_{k} = \num{0.160} \) \\
		\( m_{K^{\pm}} = \num{0.493677}\pm0.013 \) & \( f_{\phi}^{T} = \num{0.200} \) & \( m_{W} = \num{80.3692}\pm0.0133 \) & \( f_{\rho} = \num{0.209} \) \\
		\( m_{\phi} = \num{1.019431}\pm0.016 \) & \( f_{\pi} = \num{0.131} \) & \( m_{\pi^\pm} = \num{0.13957}\pm0.00017 \) & \( f_{\rho}^{T} = \num{0.165} \) \\
		\( m_{\omega} = \num{0.78266}\pm0.13 \) & \( f_{D} = \num{0.2067}\pm0.0089 \) & \( \Gamma_{\rho} = \num{0.15} \) & \( f_{B_{c}} = \num{0.489} \) \\
		\( m_{\rho} = \num{0.77526}\pm0.23 \) & \( f_{D_{s}} = \num{0.2575}\pm0.0061 \) & \( \Gamma_{\omega} = \num{8.49e-3} \) & $C_F$ = 4/3 \\
		\( m_{D^+} = \num{1.86966}\pm0.05 \) & \( f_{\omega}^{T} = \num{0.145} \) & \( \Gamma_{\phi} = \num{4.23e-3} \) & \( f_{\omega} = \num{0.195} \) \\
		\hline
		\hline\\
	\end{tabular*}
	\label{tab:syst_uncert}
\end{table}


\end{spacing}
\end{document}